\documentclass[10pt,notitlepage,a4paper]{article}

\usepackage{a4wide}
\usepackage{graphicx}
\usepackage{listings}
\usepackage{url}
\usepackage{caption}
\usepackage{comment}
\usepackage{authblk}

\usepackage{amsmath}
\usepackage{amsfonts}
\usepackage{subfig}
\usepackage{ifpdf}

\usepackage{xcolor}
\ifpdf
  \usepackage[pdftex]{hyperref}
\else
  \usepackage[breaklinks=true,ps2pdf,dvips, pdfborder={1 1 0}]{hyperref}
  \usepackage[hyphenbreaks]{breakurl}
\fi  
\definecolor{hrefcolor}{rgb}{0.0,0.0,0.8}
\newcommand{\linkcolor}{hrefcolor}

\hypersetup{pdfstartview=FitR}
\hypersetup{pdftitle={The DUNE-ALUGrid Module},
            pdfauthor={M. Alkaemper, A. Dedner, R. Kloefkorn, M. Nolte},
            breaklinks=true,
            linkbordercolor=\linkcolor,
            urlbordercolor=\linkcolor,
            citebordercolor=\linkcolor,
            runbordercolor=\linkcolor,
            menubordercolor=\linkcolor,
            filebordercolor=\linkcolor,
            runcolor=\linkcolor
            baseurl={http://users.dune-project.org/projects/dune-alugrid}
}
\hypersetup{pdfborderstyle={/S/U/W 1}}

\usepackage{xspace}
\usepackage{listings}
\newcommand{\dune}[1][]{\textsc{Dune}\ifx&#1&\else\textsc{-{#1}}\fi\xspace}

\newcommand{\alugrid}{\textsc{ALUGrid}\xspace}
\newcommand{\metis}{\textsc{Metis}\xspace}
\newcommand{\parmetis}{\textsc{ParMetis}\xspace}
\newcommand{\zoltan}{\textsc{Zoltan}\xspace}
\newcommand{\zlib}{\textsc{zlib}\xspace}
\newcommand{\dlmalloc}{\textsc{dlmalloc}\xspace}

\newcommand{\ug}{\textsc{UG}\xspace}
\newcommand{\sionlib}{\textsc{SIONlib}\xspace}
\newcommand{\alberta}{\textsc{ALBERTA}\xspace}

\newcommand{\code}[1]{ \lstinline[basicstyle=\small\sffamily]{#1} }
\newcommand{\file}[1]{\code{#1}}

\newcommand{\elem}{E}
\newcommand{\ics}{\boldsymbol{x}}

    \makeatletter
    \renewcommand\lstinline[1][]{%
        \leavevmode\bgroup 
          \def\lst@boxpos{b}%
          \lsthk@PreSet\lstset{flexiblecolumns,#1}%
          \lsthk@TextStyle
          \ifnum`}=\z@\fi
          \@ifnextchar\bgroup{%
            \ifnum`{=\z@}\fi%
            \afterassignment\lst@InlineG \let\@let@token}{%
            \ifnum`{=\z@}\fi\lstinline@}}
    \makeatother

\newcommand{\tbl}[2]{\begin{center}\caption{#1}{#2}\end{center}}
\newcommand{\setR}{\mathbb{R}}
\newcommand{\setN}{\mathbb{N}}
\newcommand{\identity}{\mathbb{I}}

\begin{document}

\title{The \dune[ALUGrid] Module}
\author[a]{Martin Alk\"amper}
\author[b]{Andreas Dedner}
\author[c]{Robert Kl\"ofkorn}
\author[d]{Martin Nolte}
\affil[a]{University of Stuttgart}
\affil[b]{University of Warwick}
\affil[c]{International Research Institute of Stavanger}
\affil[d]{University of Freiburg}



\providecommand{\keywords}[1]{\textbf{Keywords: } #1}


\date{August, 2015}

\maketitle

\begin{abstract}
In this paper we present the new \dune[ALUGrid] module. This module
contains a major overhaul of the sources from the \alugrid library and the
binding to the \dune software framework. The main improvements concern the
parallel feature set of the library. The main changes include user-defined load
balancing, parallel grid construction, and an redesign of the 2d grid which can now
also be used for parallel computations. In addition many improvements
have been introduced into the code to increase the parallel efficiency and
to decrease the memory footprint. 

The original \alugrid library is widely used within the \dune
community due to its good parallel performance for problems requiring local
adaptivity and dynamic load balancing. Therefore, 
this new model will benefit a number of \dune users. In addition we have
added features to increase the range of problems for which the grid manager can
be used, for example, introducing a 3d tetrahedral grid using a parallel 
newest vertex bisection algorithm for conforming grid refinement.
In this paper we will discuss the new features, extensions to the \dune
interface,  and explain for various 
examples how the code is used in parallel environments. \\ 

\noindent
\end{abstract}

\section{Introduction}
The origin of the \alugrid package dates back to the late 20th century 
as part of the PhD thesis of Bernhard Schupp 
\cite{schupp:phd}.
Back then, the task was to develop a software that could solve the
compressible Euler equations of gas dynamics with a Finite Volume scheme 
on a parallel computer in 3d including local grid adaptivity.
To achieve this task Schupp implemented a 3d hexahedral adaptive mesh
including dynamic load balancing based on METIS graph partitioning \cite{metis}.
Later, support for tetrahedral elements was
added\footnote{Tetrahedral element support was implemented by M. Ohlberger.} to the grid manager 
and the code was successfully used to simulate 
solar eruption phenomena based on the MHD equations \cite{dedner:MHDCode}.
Shortly after this, the library was used to implement the \dune grid interface 
\cite{dunepara:05}. The \alugrid bindings were the first grid
implementation providing the full interface for an adaptive,
distributed grid including dynamic load balancing.
It has been shown that \alugrid is a very efficient
implementation of the \dune grid interface. For an explicit Finite Volume scheme, the performance loss
introduced by the \dune bindings is roughly $10\%$ compared to the native \alugrid
implementation~\cite{dunepaperII:08,dunepara:05,kloefkorn:phd}. 
At that time also a serial 2d simplex grid was added to the code basis.
The following releases of the software saw only maintenance work with no
substantial increase in the feature set.

In this paper the first major overhaul of the \alugrid code basis is
described. Originally, \alugrid was available as a stand alone library with a quite complex user API.
Consequently, \alugrid was used exclusively through the bindings available
in the \dune[Grid] module. To reduce maintenance and improve usability we integrated
both, the original \alugrid library and its bindings to \dune, into the new
\dune[ALUGrid] module. 
In addition a number of new features have been added and the
efficiency of the code has been increased while the memory footprint has
been substantially reduced.

\alugrid is a capable and reliable parallel-adaptive grid manager and 
has been used in codes based on \dune, for example, in life science
applications \cite{kopf:13,jehl:14}, in the simulation of nanotechnology \cite{may:09,Fallahi:12}, 
in simulations related to numerical weather and climate prediction \cite{dunecosmo:12,mueller:14}, 
simulation of reactive flow in a moving domain \cite{motor:13}, or in 
subsurface simulations \cite{faigle:14} and other works.

Within \dune the other unstructured grid manager capable of parallel-adaptive 
computations is \ug~\cite{lang:05} with the \code{UGGrid} realization of the \dune grid
interface. 
A comparison for time-explicit applications with and without adaptivity 
using the different grid implementations available 
in \dune is presented in~\cite{perfpit:12}. 

Besides a vast number of structured or Cartesian grid managers supporting adaptive
refinement (see \url{http://math.boisestate.edu/~calhoun/www_personal/research/amr_software/})
there exist a few other open source unstructured grid managers (at present
without bindings to \dune), for example, 
deal.II~\cite{dealII81} which is build on top of p4est~\cite{burstedde:11}. Hexahedral
grids with non-conforming refinement are provided. Excellent scalability has been reported.
As a drawback, the macro mesh has be present on every core limiting the macro mesh size.
Other very capable unstructured grid managers are, for
example, the "Flexible Distributed Mesh Database (FMDB)"~\cite{fmdb:12},
\texttt{libMesh}~\cite{libMeshPaper}, or AMDIS~\cite{amdis}. The latter is providing
tetrahedral elements with bisection refinement. 


In this paper we present work done in recent years to improve the useablility, efficiency, and reduce 
maintanace cost of \alugrid:

In the previous versions of \alugrid the implementation of the 2d and 3d grids
were completely seperate. This resulted in a disjoint set of features with the 2d grid
implementing bisection not available for the 3d grid while at the same time the 2d
grid did not provide any parallel features. In \dune[ALUGrid] the original code for the
2d grid has been removed. Grids in two space dimensions or surface grids are now 
implemented by embedding them into three space dimensions, making it possible to
directly use the 3d grid implementation. The main advantage of this is the significant 
reduction in code maintance while at the same time all improvements in performance or
feature set of the 3d code will be directly available also for 2d grids.
Furthermore, since conforming bisection is now also available in 3d, this merge has not
resulted in any loss of functionality. 

To simplify the installation, the \dune bindings and the library itself have been combined in a single \dune module. 
This module includes a number of new features,
which make the \alugrid implementation a lot more flexible and make it
possible to use it through \dune for a wider range of problems:
\begin{itemize}
\item {\bf extension to implement a wider range of methods:} \\
      the main extension is conforming grid refinement implemented in the
      parallel 3d code. Furthermore the 2d grid can be used for distributed computations
      so that the 2d and 3d code now share the same feature set.
      In addition the support for quadrilateral and surface grids in 2d and 
      periodic boundary treatment in 3d for parallel computations has been
      improved.
\item {\bf increasing usability and efficiency:} \\
      the memory footprint is considerably reduced (Section~\ref{sec:memory}),
      a cleaner interface for callback adaptation, which was partially
      available before, is discussed in Section~\ref{sec:adaptcallback}.
\item {\bf increasing usability and efficiency for parallel computation:} \\
      new features include:
      parallel grid construction (discussed in Section~\ref{sec:parallelgrid}),
      backup and restore (discussed in Section~\ref{sec:dataio}),
      overlapping communication and computation, (discussed in 
      Section~\ref{sec:communication}),
      wider range of load balancing algorithms 
      (including internal implementations), and user-defined partitioning algorithms
      (these are discussed in Sections~\ref{sec:userdeflb} and \ref{userdeflb}).
      In summary \dune[ALUGrid] now provides
  \begin{itemize}
    \item intrinsic partitioning based on space filling curves, making
    \dune[ALUGrid] independent of other packages, 
    \item bindings for the library \zoltan \cite{zoltan},
    \item an interface for user-defined partitioning.
  \end{itemize}
\end{itemize}

In Section~\ref{sec:performance testing} we describe how we have evaluated the
performance of the \dune[ALUGrid] module and report on a number of different strong and
weak scaling results obtained on both a computing cluster and a highly integrated
high performance computing system. Following, in Section~\ref{sec:using dune-alugrid}, we present the new features
and interface extensions from a user's point of view, 
Finally we make some concluding remarks and discuss some open issues
with this module. 

For a successful and satisfactory reading of this paper it is 
recommended to be familiar with the \dune papers \cite{dunepaperII:08,dunepaperI:08}
and the terminology used therein. 

\section{Performance Testing}
\label{sec:performance testing}

The aim of \cite{schupp:phd} was to develop an efficient parallel
implementation of an adaptive explicit Finite Volume scheme. These schemes are
widely used for solving hyperbolic conservation laws. The appearance of steep
gradients or shocks in the solution make grid adaptivity a good choice to
increase the efficiency of a scheme. These shocks move in time requiring the
refinement zones to move with the shocks and coarsening to take place
behind them. In combination with a domain decomposition approach for
parallel computation, this means that the load is difficult to balance
between processors and dynamic load balancing is essential. So in each time
step the grid needs to be locally refined or coarsened and the grid has to be
repartitioned quite often. What makes this problem extremely challenging is
the fact that evolving the solution from one time step to the next is very
cheap since the update is explicit and no expensive linear systems have 
to be solved. So adaptivity and load balancing will dominate the
computational cost of the solver. Both of these steps require global
communication steps and the communication of possibly a significant amount
of data and are therefore difficult to implement even with a moderate
amount of parallel efficiency (see for example \cite{burstedde:11}).

Since this is a very demanding problem for a parallel code,
we have decided to continue using explicit Finite Volume schemes to measure the
performance of the \dune[ALUGrid] module. Grid performance plays a role in 
matrix-free methods where frequent grid iteration occurs 
in order to evaluate differential operators even if the used discrete function space is of higher
order.
In contrast, the performance of implicit matrix-based methods will have a
stronger dependency on the efficiency of the parallel solver package
than on the grid implementation.
Therefore, testing implicit methods would not provide as much insight into the
performance of the grid module itself.

As a simple example, we consider the scalar transport equation
\begin{equation*}
\partial_t u + \nabla\cdot \bigl( (1.25,1.25,0)^T\,u \bigr) = 0
\end{equation*}
with suitable initial and boundary data (see
\file{examples/problem-transport.hh}).
In the adaptive Finite Volume scheme we use an upwind numerical flux and a
jump indicator to trigger grid adaptation.

For a more damanding example, we also apply this scheme to the Euler equations
of gas dynamics
\begin{equation*}
  \partial_t \begin{pmatrix} \rho \\ \rho\,\vec{v} \\ \epsilon \end{pmatrix}
    + \nabla \cdot \begin{pmatrix} \rho\,\vec{v} \\ \rho\,\vec{v} \otimes \vec{v} + p\,\identity \\ (\epsilon + p)\,\vec{v} \end{pmatrix}
    = 0,
\end{equation*}
where $\identity \in \setR^{d \times d}$ denotes the identity matrix.
We consider an ideal gas, i.e.,
$p = (\gamma - 1)\,(\epsilon - \frac{1}{2}\,\rho\,\lvert \vec{v} \rvert^2)$,
with the adiabatic constant $\gamma = 1.4$.
In the adaptive scheme, we use an HLLC numerical flux \cite{toro:09} in the
evolution step and the relative jump in the density to drive the grid adaptation.
Two typical test problems found in the literature, the Forward Facing Step and
the interaction between a shock and a bubble (see \cite{limiter:11} and
references therein) are implemented (see \file{examples/problem-euler.hh}).

To benchmark solely adaptation and load balancing, we implemented a third, even
more demanding test case.
Instead of using the solution to a partial differential equation to determine
the zones for grid refinement and coarsening, a simple boolean function
$\elem \mapsto \eta_{\elem}$ is used (see \file{examples/problem-ball.hh}).
We refine all elements located near the surface of a ball rotating around the
center of the 3d unit cube: 
\begin{equation}
  \label{test:ball}
  \begin{aligned}
    \boldsymbol{y}( t ) &:= \Bigl( \tfrac{1}{2} + \tfrac{1}{3} \cos(2\pi t), \tfrac{1}{2} + \tfrac{1}{3} \sin(2\pi t), \tfrac{1}{2} \Bigr)^T, \\
    \eta_{\elem} &:=
    \begin{cases} 
      1  &  \text{if } 0.15 < |\ics_{\elem} - \boldsymbol{y}( t )| < 0.25, \\ 
      0  &  \text{otherwise,}
    \end{cases}
  \end{aligned}
\end{equation}
where $\ics_{\elem}$ denotes the barycenter of the element $\elem$.
A cell $E$ is marked for refinement, if $\eta_{\elem} = 1$ and for coarsening otherwise.
This sort of problem was also studied in \cite{schupp:phd}.
Since the center of the ball is rotating, frequent refinement and coarsening occurs,
making this an excellent test for the implemented adaptation and load balancing
strategies.

\subsection{Memory Consumption}
\label{sec:memory}
Memory consumption has become more and more critical for any numerical software 
since the overall memory available per core has declined lately. 
First we need to give a short summary of the data structure used to store
grid elements:
A vertex stores its coordinates, an edge stores pointers
to the two vertices, a quadrilateral face stores pointers to the four edges and a hexahedron stores
pointers to the six faces it consists of. For example, the memory consumption of a vertex on a
64bit architecture is $56$ bytes due to the storage of coordinates ($3$ \code{double} result in $24$ bytes), 
$8$ bytes for the vtable (all interfaces in \alugrid use dynamic polymorphism),
$8$ bytes for a pointer to the grid class, 
and another $12$ bytes for flags, reference counting, and index storage, which due to padding this adds up to
$56$ bytes.

\begin{table}[ht]{}
\renewcommand{\arraystretch}{1.5}
\tbl{Memory consumption by ALUGrid's entities in bytes (in braces we put the memory
consumption in  ALUGrid's 1.52 version).}{%
\begin{tabular}{l|rr|rr}
  type         &   \multicolumn{1}{c}{tetra}  & \multicolumn{1}{c|}{ hexa} & \multicolumn{1}{c}{  macro tetra} & \multicolumn{1}{c}{ macro hexa} \\ \hline \hline 
  vertex       &   56 \phantom{5}(64)  &  56 \phantom{5}(64)  &   80 \phantom{5}(80)   &  80 \phantom{5}(80)  \\
  edge         &   56 (136) &  56  (136) &   64  (144)  &  64  (144) \\ 
  face         &   88 (160) &  96  (174) &   96  (168)  &  104 (184) \\ 
  element      &   96 (160) &  112 (184) &   104 (168)  &  120 (192) \\  
\end{tabular}}
\end{table}

\begin{figure}
  \subfloat[cube]{
  \includegraphics[width=0.45\linewidth]{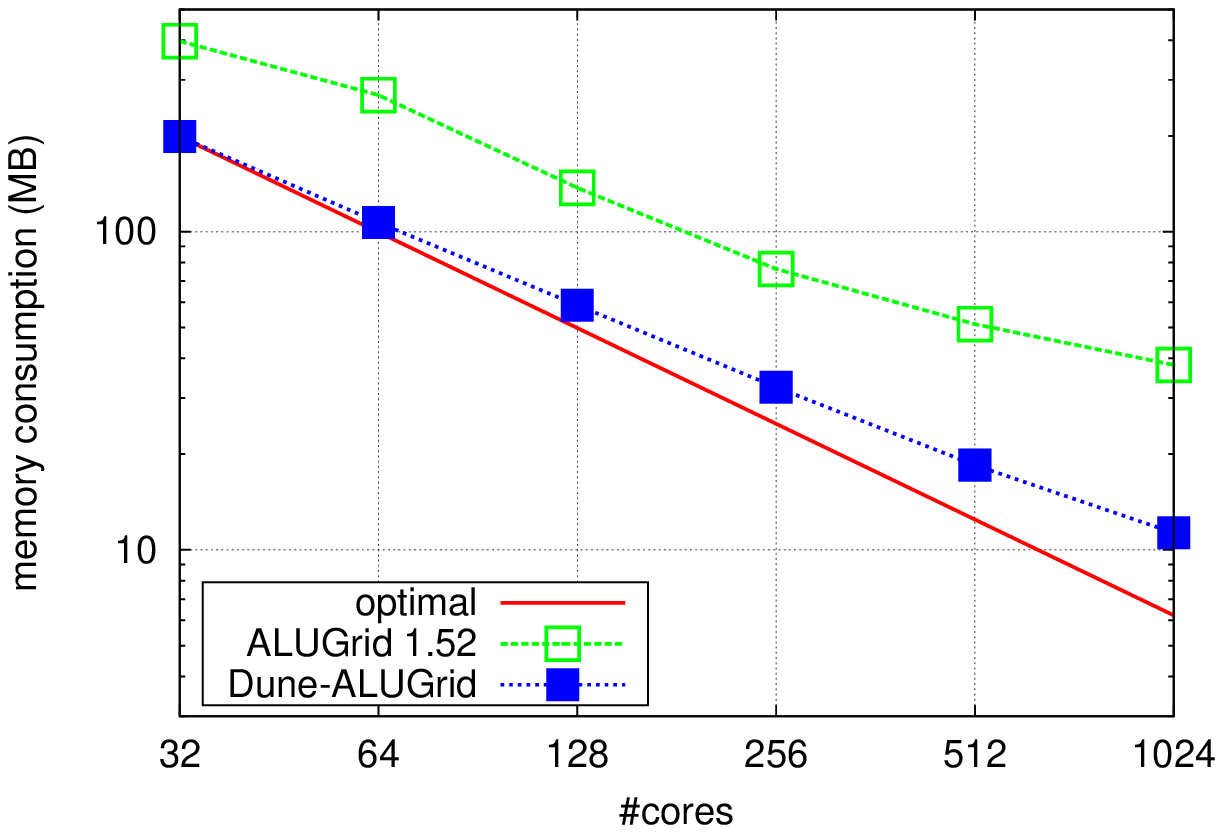}}
  \subfloat[simplex]{
  \includegraphics[width=0.45\linewidth]{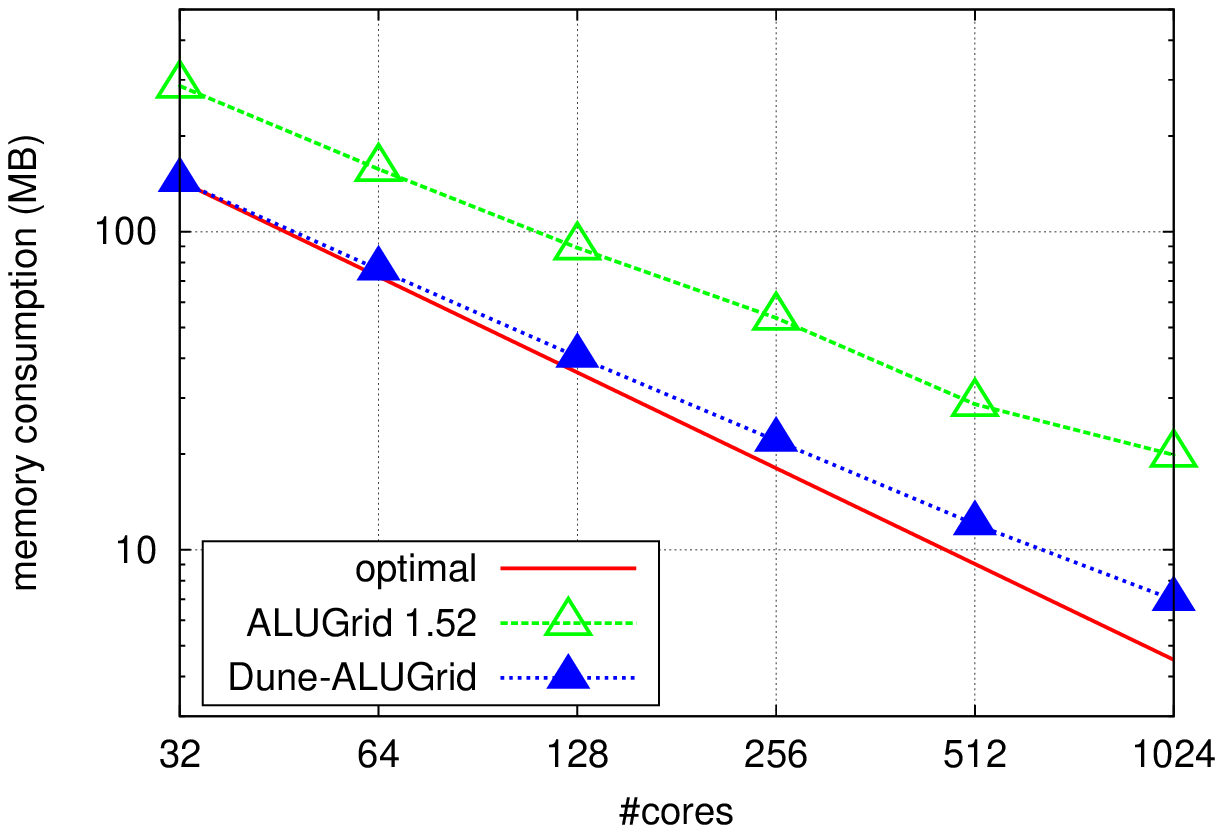}}
  \caption{Comparison of memory usage for the old and the new version. Both version use
           dlmalloc as memory allocator. The $1.52$ version has been patched for this purpose.}
  \label{fig:memcompare}
\end{figure}

Depending on the size of the macro grid and the face/edge to element ratio a hexahedral
grid consumes between $700$ and $800$ bytes per element. The tetrahedral version of the grid 
consumes between $350$ and $400$ bytes per element. Note that these numbers strongly
depend on the macro grid chosen and might vary for other macro grids. 
For the old version $1.52$ storing a hexahedral element needed between $1\,300$ and
$1\,500$ bytes. For a tetrahedral element version $1.52$ needed between $650$ and $750$ bytes.  
In Figure \ref{fig:memcompare} we show the memory consumption for the old and the new
version for the ball test case with adaptation using the refinement from \eqref{test:ball}. 
In summary the memory consumption has been reduced by about a factor of $2$.

In an adaptive grid, entities are frequently created during refinement and
destroyed during coarsening.
As \alugrid allocates memory for each grid entity separately, efficient memory
allocation and deallocation plays an important part in this process.
To allow for customization, \alugrid derives all entities from an object
called \code{MyAlloc}, which contains overloaded operators \code{new} and \code{delete}.
Two such objects are shipped with \dune[ALUGrid]. 
\begin{description}
\item[default]
    does not overload the operators \code{new} and \code{delete}, so that
    standard C++ memory allocation is used. This is the default memory allocation used.
\item[dlmalloc]
    makes use of Doug Lea's memory allocator (\code{dlmalloc}) \cite{dlmalloc:96},
    which can be downloaded from \url{http://g.oswego.edu/dl/html/malloc.html}.
    If the configure option \\ 
    \texttt{--with-dlmalloc=PATH} 
    is provided specifying a path to the \code{dlmalloc} installation, 
    \code{dlmalloc} will be used for allocation of grid entities.
\end{description}

\begin{figure}[!ht]
\begin{center}
  \includegraphics[width=0.65\linewidth]{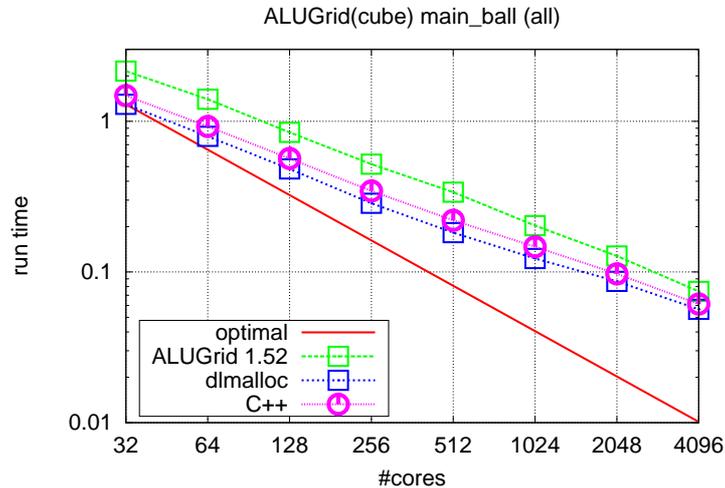}
\end{center}  
  \caption{Comparison of run times for the different memory allocation strategies. The
    memory allocation using Doug Lea's memory allocator \cite{dlmalloc:96}
    performed best, the strategy used in \alugrid 1.52 performed worst and has
    therefore been removed in the new version. 
    For load balancing we used the internal space filling curve approach with locally
    computed linkage (partition method id 4).}
  \label{fig:memman}
\end{figure}

In Figure \ref{fig:memman} we present a comparison of runtimes between the different
memory allocation strategies. 
The former internal \alugrid implementation based on \code{std::map} and \code{std::stack} 
has been removed since it did not lead to performance gains, anymore.
For adaptation with the ball refinement from equation \eqref{test:ball} using
\code{dlmalloc} around $10\,\%$ less CPU time is consumed in comparison to the standard
C++ memory allocation on Yellowstone \cite{Yellowstone}.

\begin{figure}
  \subfloat[memory]{\label{fig:memcompare2d_mem}
  \includegraphics[width=0.45\linewidth]{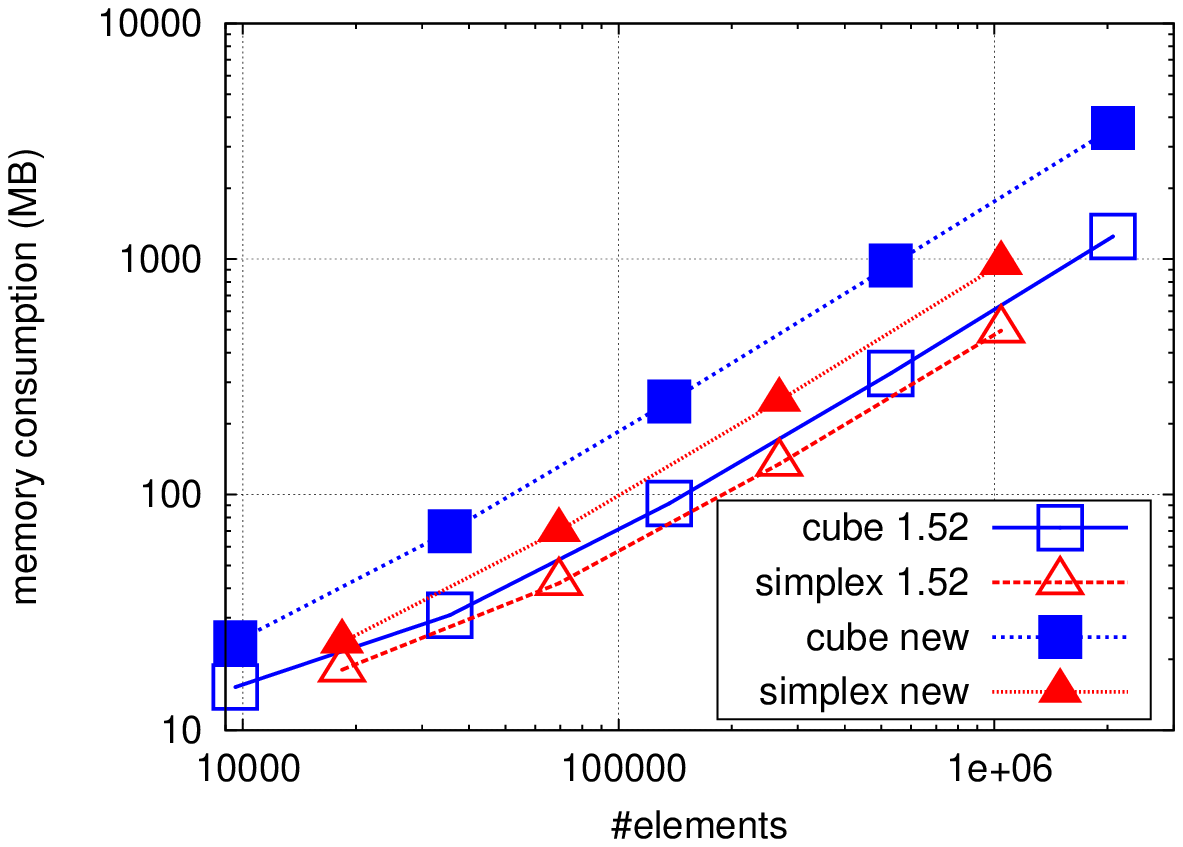}}
  \subfloat[run times]{\label{fig:memcompare2d_run}
  \includegraphics[width=0.45\linewidth]{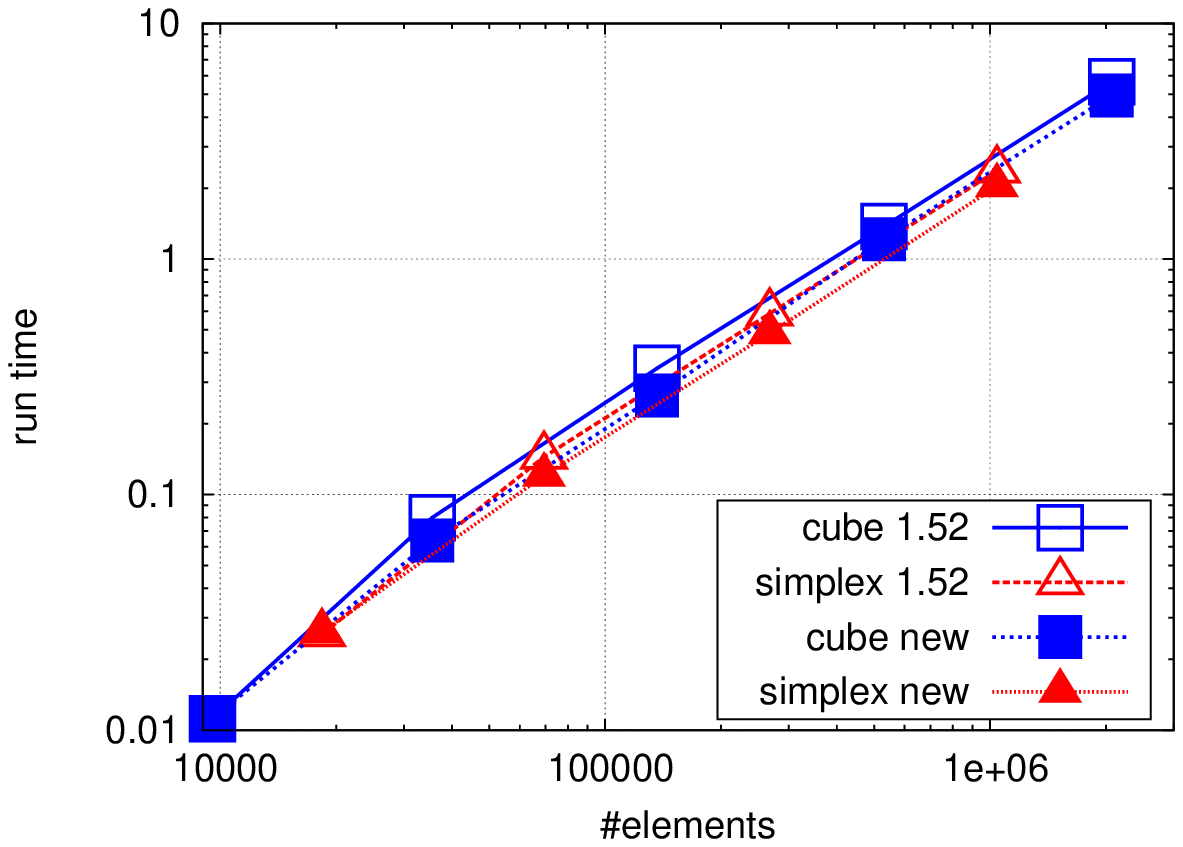}}
  \caption{Comparison of memory usage and run times for the 2d version
           in the old and the new implementation. Both versions use
           dlmalloc as memory allocator. The $1.52$ version has been patched for this purpose.}
  \label{fig:memcompare2d}
\end{figure}

As mentioned in the introduction the codes for the 2d and 3d grid have been unified. 
The only drawback of embedding the 2d into a 3d
grid is an increase in the memory requirements of the 2d grid. For example a 2d
quadrilateral grid is modelled using a 3d hexahedral grid by replacing each quadrilateral by one
hexahedron. Effectively this leads to a doubling of the memory usage in this case.
For a triangular grid the resulting increase in memory usage is less severe.
This is confirmed by the results shown in Figure~\ref{fig:memcompare2d_mem}.
This increase in memory consumption in the new version in compensated by
improvements in performance, as can be seen in Figure~\ref{fig:memcompare2d_run}.

\subsection{Scaling results}
\begin{figure}[!ht]
\begin{center}
  \includegraphics[width=0.75\linewidth]{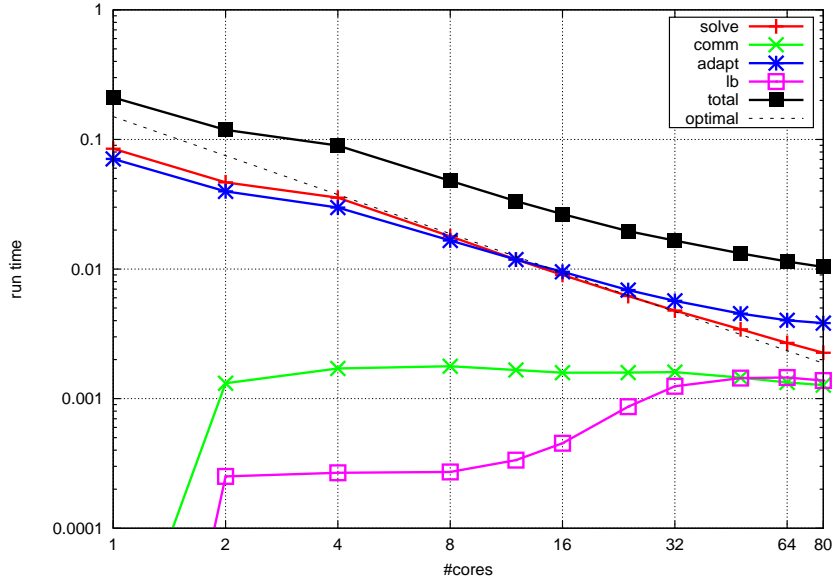}
  \caption{Strong scaling results for 2d Euler shock-interaction problem on a small size
  computer cluster. 
  }
  \label{fig:speedup_strong_cluster2d}
\end{center}
\end{figure}

We start with testing the new parallel version of the 2d code.
In Figure \ref{fig:speedup_strong_cluster2d} we present the results of a 2d
version of the shock-bubble interaction problem taken from \cite{limiter:11} 
using a small size computer cluster consisting of 20 Intel Core-i3 2100 (Sandy-Bridge)
desktop computers connected via standard gigabit ethernet.
The curves represent different parts of the code
(\emph{solve}:\ computation and synchronization of the update vector,
\emph{comm}:\ global synchronization of time step,
\emph{adapt}:\ grid adaptation, and
\emph{lb}:\ load balancing).
Finally we show the \emph{total} runtime. Note that there are some small
parts of the time loop not seperately shown so that the total runtime is
not exactly the sum of the four parts shown. 
For the dynamic load
balancing we use the space filling curve approach newly implemented in
\dune[ALUGrid], see Section~\ref{sec:userdeflb}.
The grid load balancing is checked every $25^{th}$ time step and 
is performed when the number of elements between the largest
and smallest partition differs by $20\%$ or more. 

As we can see the Finite Volume part of the 
code scales very well up to $64$ cores. The overall scaling is still
acceptable. Note that we are using hyperthreading to execute four
processes per node although these are dualcore machines. Parallel
efficiency increases by about $10\%$ when only two proceses are put on one
node but the runtime using a given number of nodes is quite a bit higher. 
Note that the number of degrees of freedom 
was quite small in this simulation so that even on a
few cores, the cost for the solve step and for the adaptation are comparable. Thus the
total runtime more or less follows the curve for the adaptation cost leading to
$50\%$ efficiency going from $4$ to $64$ cores while the solve step itself is still
close to optimal.

\begin{figure}[!ht]
\begin{center}
  \includegraphics[width=0.75\linewidth]{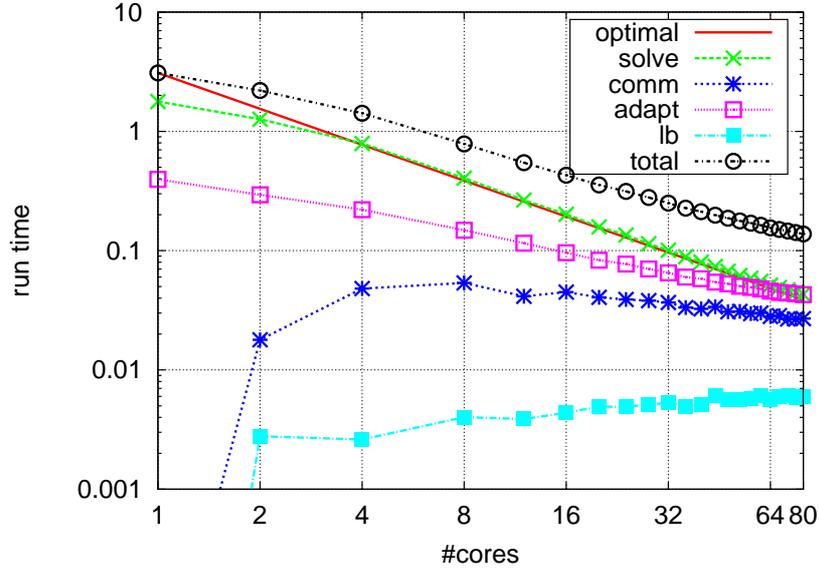}
  \caption{Strong scaling results for 3d Euler shock-interaction problem on a small size
  computer cluster. The macro grid contains $4\,096$ hexahedrons which is also the
  coarsest grid and the maximal refinement level is set to $4$ (parameter $22\ 0\ 4$).}
  \label{fig:speedup_strong_cluster3d}
\end{center}
\end{figure}

We repeated the same test but now using the 3d grid (Figure \ref{fig:speedup_strong_cluster3d}). 
The macro grid was larger in this
simulation and combined with a slightly higher per element
cost of the 3d Finite Volume scheme, the solve step dominates the adaptation
up to 64 cores. The efficiency going from $4$ to $64$ cores is thus higher with an
value at about $70\%$.

\begin{figure}[!ht]
\begin{center}
  \includegraphics[width=0.75\linewidth]{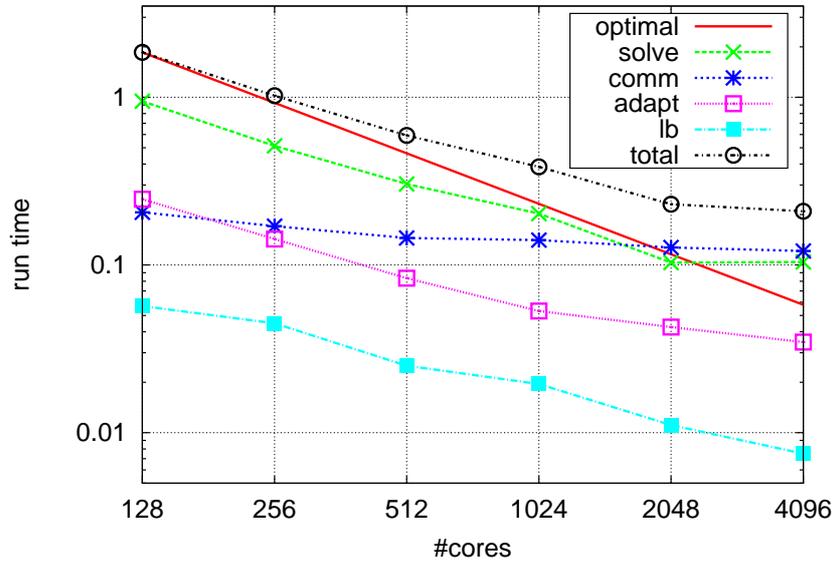}
  \caption{Strong scaling results for Euler shock-interaction problem on the peta scale
  supercomputer Yellowstone \cite{Yellowstone}. 
  The macro grid contains $32\,768$ hexahedrons which is also the 
    coarsest grid and the maximal refinement level is set to $6$ (parameter $23\ 0\ 6$).}
  \label{fig:speedup_strong_hpc}
\end{center}
\end{figure}

In Figure \ref{fig:speedup_strong_hpc} we present resuts for the same
computation but this time on the \emph{Yellowstone} supercomputer
\cite{Yellowstone}. We made two
changes to the settings described above which increase performance on large
core counts with a strong interconnect: we use the space filling curve
approach with linkage storage (see Section~\ref{sec:userdeflb}) and instead of
rebalancing when the partitions differ by $20\%$, the grid is repartitioned
already when the inbalance is more than $5\%$.
Efficiency is quite good up to $2\,048$ processors but after that the problem
size is to small to adequately distribute among $4\,096$ core 
and no noticeable performance increase is achieved. At
this point the communication cost becomes comparable to the actual
evolution step. The grid adaptation stage is still scaling well at $1000$
cores while the loadbalancing starts becoming less efficient earlier. But
the computational costs of these two parts of the algorithm is still quite 
small compared to the evolution step. Note that in the previous cluster case with
its slow interconnect the loadbalancing step was not scaling at all. 

\begin{figure}[!ht]
\begin{center}
  \includegraphics[width=0.75\linewidth]{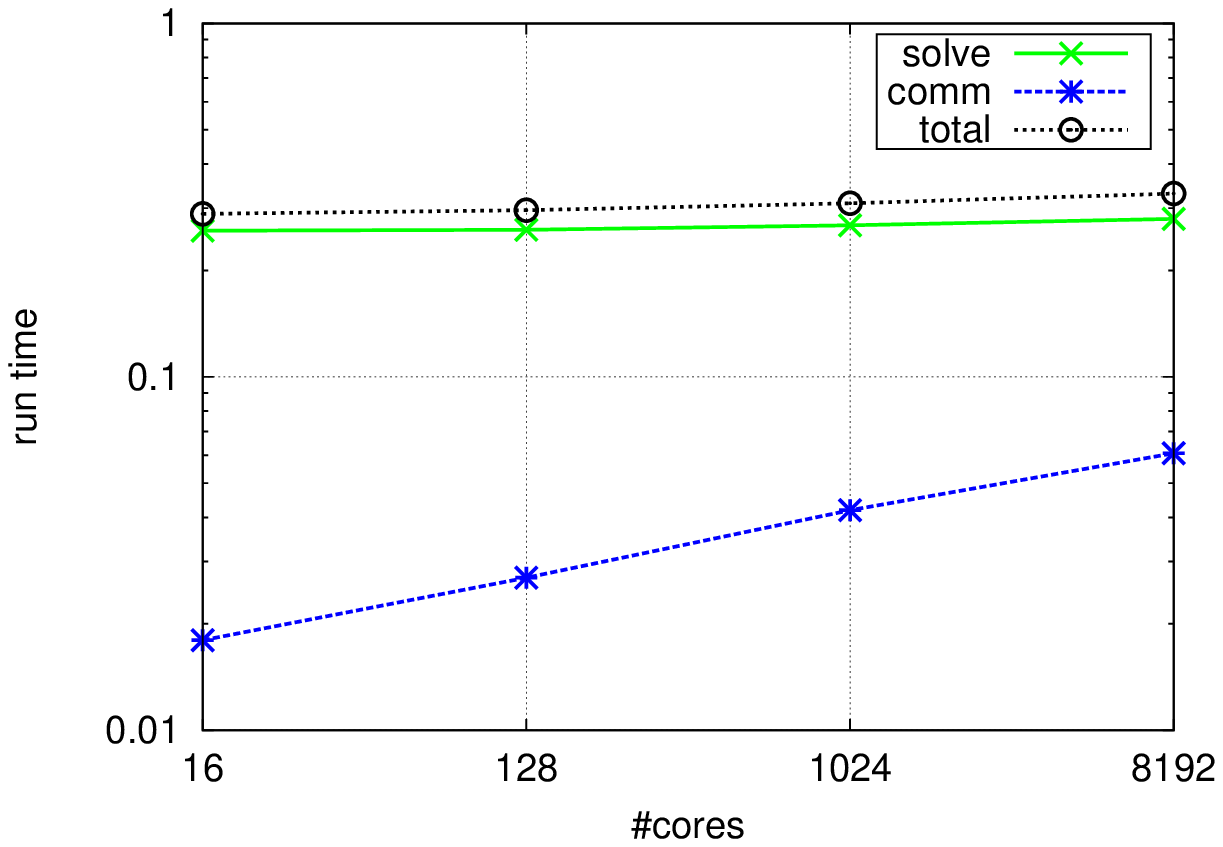}
  \caption{Weak scaling results for Euler shock-interaction problem on the peta scale
    supercomputer Yellowstone \cite{Yellowstone}. The number of elements is kept
    constant per core at $131\,072$ hexahedrons (parameter $25\ 2\ 0$).}
  \label{fig:speedup_weak_hpc}
\end{center}
\end{figure}

The computations reported on above were strong scaling tests, keeping the
problem the same and only increasing the number of cores used. Thus the
computational cost is reduced while increasing the parallelization overhead at the
same time. In addition parallel efficiency is difficult to achieve since 
obtaining a good load distribution becomes challenging when the problem size is fixed. 
Therefore, we also include a weak scaling test in Figure \ref{fig:speedup_weak_hpc}.
Since with adaptive simulations it is difficult to increase the  problem size
in the systematic way necessary for weak scaling experiments, we have
performed a fixed grid computation here.
As can be seen the
computational cost only slowly increases leading to high parallel
efficiency of $88\%$ going from $16$ to $8192$ cores. 

\section{Using the \textsc{DUNE-ALUGrid} Module}
\label{sec:using dune-alugrid}

This section discusses the features of \dune[ALUGrid] from a user perspective.
Special emphasis will be put on extensions to the \dune grid interface.

\subsection{Structure of the Module}
The structure of the new module is as follows:
the main code for the grid implementation and the \dune bindings are in the
\file{dune} folder of the \dune[ALUGrid] module. A program to read in a macro grid on a single
processor and to write a partitioned version in a binary format to a file is
provided in the \file{utility} folder. Finally the \file{examples} folder
contains the main executables for testing the \dune[ALUGrid] modules. 
All the test problems can be used with any grid manager implementing the \dune[Grid] interface.
This makes it not only possible to test the \alugrid implementation but also
to compare with other realizations of the \dune grid interface.
The code is very similar to the example provided in the \dune[Fem-Howto] and
comparable with the tutorial found in the \dune[Grid-Howto].

The code is mainly distributed across four files;
\begin{description}
\item[main.cc]  contains the initial grid construction and the time loop.
\item[fvscheme.hh] contains the computation of the update vector
  and the marking strategy.
\item[adaptation.hh] contains the code for carrying out the grid modification.
\item[piecewisefunction.hh] contains all classes used to handle the degrees of
  freedom including storage, restriction and prolongation, and
  communication.
\end{description}

Switching between the three different test cases is done via pre-processor
flags or by making one of the three executables
\file{main_ball}, \file{main_transport}, or \file{main_euler}.
Each program takes three command line parameters:
\begin{verbatim}
  ./main [problem-nr] [startLevel] [maxLevel]
\end{verbatim}
The first one determines the test case to use (including initial data and macro grid);
\code{startLevel} and \code{maxLevel} determine the coarsest and finest grid level,
respectively.


Extensions of the \dune grid interface discussed in
this paper can be tested in different sub-folders of \file{examples}.
The basic code is always the same with the necessary changes described
in detail in the following chapters.
There are four sub-folders, each containing a script, to compare the original and the modified
implementation. Again, pre-processor defines are used to provide different
implementations in the same code:

\begin{description}
\item[callback]
      compare dof storage and callback adaptation in serial.\\
      Script: \file{check-adaptation.sh}\\
      Pre-processor flags: \code{CALLBACK} and \code{USE_VECTOR_FOR_PWF}.
\item[communication]
      test asynchronous communication with callback adaptation and persistent
      container (best from before)\\
      Script: \file{check-communication.sh} \\
      Pre-processor flags: \code{NON_BLOCKING}.
\item[loadbalance]
     test the extensions to the loadbalancing interface. In addition to the internal
     loadbalancing methods, user-defined weights can be added (preprocessore flag
     \code{USE_WEIGHTS} and a simple user-defined loadbalancing strategy
     is available (flag \code{USE_SIMPLELB}. With the flag \code{USE_ZOLTAN} a complete
     reimplementation of the internal zoltan bindings is available based on the
     extensions of the grid interface (requires the configure option
     \code{enable-experimental-grid-extensions}).
\item[testEfficiency]
      test on one computer using multi-core, e.g.\ $1\to2\to4\to8$, and
      test on cluster with N computers and P cores, e.g., $P\to2P\to4P\to8P$.
      By changing the pre-processor flags in the script different versions
      can be tested.\\
      Script: \file{check-efficiency.sh}\\
      Pre-processor flags: \code{CALLBACK}, \code{USE_VECTOR_FOR_PWF}, \code{NON_BLOCKING}.
\end{description}

Note that by default the cube version of \dune[ALUGrid] is used.
This can be changed in \file{Makefile.am}.

\subsection{Configuration}
The new \dune[ALUGrid] module is available via the module home page 
\url{http://users.dune-project.org/projects/dune-alugrid}. 
The repository can be accessed using the  \code{git} repository from
\url{https://users.dune-project.org/repositories/projects/dune-alugrid.git}.

The \dune[ALUGrid] module depends on \dune[Grid] and can be easily
configured using the \dune build system. Using \dune[ALUGrid] in a user module then
only requires adding a dependency (or suggestion) in the \file{dune.module} file,
including \file{dune/alugrid/grid.hh}, and using
\begin{center}
\code{Dune::ALUGrid< dimgrid, dimworld, eltype, refinetype, communicator >}
\end{center}
with $2 \le$ \code{dimgrid} $\le$ \code{dimworld} $\le 3$ for grid and world dimension,
\code{eltype} $=$ \code{Dune::simplex},\code{Dune::cube}, and
\code{refinetype} $=$ \code{Dune::conforming},\code{Dune::nonconforming}.
In this version, the only restriction is that conforming refinement is
not a valid choice for cube grids.
Contrary to previous versions, conforming refinement for a 3d simplex grid is
now available.
For the communicator, either \code{ALUGridMPIComm} for a parallel grid or
\code{ALUGRIDNoComm} for serial grid can be used.
By default, MPI communication is used, if available. Note that if
\dune[ALUGrid] was compiled in parallel mode then MPI has to be initialized
before constructing a grid object even in a serial computation.

There are a number of packages which can be used to increase the flexibility and
performance of the \dune[ALUGrid] module. Paths to the installed versions of these
packages have to be provided during the configuration of the module, i.e., within the
configuration file used in the call of the \code{dunecontrol} script:

\begin{description}
\item[\tt{--with-dlmalloc=PATH}:]
path to Doug Lea's malloc library (required version $>= 2.8.6$). If this library is
available the memory management for \dune[ALUGrid] will use the \dlmalloc package
\cite{dlmalloc:96}.
This can improve performance as shown in Section~\ref{sec:memory}.

\item[\tt--with-metis=PATH:]
path to the \metis library \cite{metis}. If available, \metis can be used for
load balancing.

\item[\tt--with-metis-lib=NAME:]     
name of the metis libraries (default is \code{metis}).

\item[\tt--with-zoltan=PATH:]
path to the \zoltan package. This package provides a wide range of additional
load balancing methods including those provided by \metis and \parmetis.
Details on how to use different load balancing methods are provided in
Section~\ref{sec:userdeflb}.

\item[\tt--with-zlib=PATH:]
path to \zlib \cite{zlib}. If available, \zlib compression can be used for
backup and restore of a full \dune[ALUGrid] grid object.
More details on data I/O are provided in Section~\ref{sec:dataio}.
\end{description}

\subsection{Parallel Grid Construction}
\label{sec:parallelgrid}
Any grid-based numerical simulation must at some time construct a grid of the
computational domain.
The general \dune grid interface assists this step by providing three basic
construction mechanisms:
\begin{description}
\item[GridFactory]
    is a general interface for the construction of unstructured grids.
    Basically, it constructs the grid from a list of vertex coordinates and
    a list of elements.
\item[StructuredGridFactory]
    can be used to construct a grid of an axis-aligned cube domain.
    For unstructured grids, a default implementation based on the GridFactory is
    provided.
\item[GridReaders] can be used to read files given in a special format.
    These readers will generally use the \code{GridFactory} to construct the
    grid. A \dune specific format is available through the \code{DGF}
    reader. 
    An extension of this format to partitioned grids is discussed in Section~\ref{sec:paralleldgf}.
\end{description}
Additionally, \dune[ALUGrid] provides a native file format for predistributed
macro grids.

At the time of this writing, the \code{GridFactory} interface does not support
the construction of unstructured grids in parallel.
The entire grid must first be constructed on one process and then distributed to
all processes using the load balancing algorithm.
For large macro grids, this method is at least inefficient if not impossible as
the macro grid might not even fit into the memory of one computational node.
Without specialization, this restriction also holds for the
\code{StructuredGridFactory} and the DGF parser. 
\dune[ALUGrid] overcomes this difficulty by providing specializations of all
three grid construction mechanisms.
In addition the \dune[ALUGrid] module contains
utility tools to perform the distribution off line, writing native
distributed \alugrid files for use in the actual computation. These will be described at
the end of this section. 

\subsubsection{The GridFactory}

In \dune, the construction of unstructured grids is handled by the
\code{GridFactory} class, which has to be specialized for each grid
implementation supporting them.
The most important interface methods are
\begin{lstlisting}
  void insertVertex ( const Dune::FieldVector<ctype,dimensionworld> &coord );
  void insertElement( const Dune::GeometryType &type,
                      const std::vector<unsigned int> &vertices );
  void insertBoundarySegment( const std::vector<unsigned int> &vertices );
\end{lstlisting}

The main difficulty when constructing a predistributed grid is the identification of
the process boundaries. Using a large amount of global communication and coordinate
comparison this could be achieved using the interface provided by the \dune[Grid]
module. Since this is neither efficient nor very reliable, we extend the interface
requiring the user to provide a globally unique number for each vertex in the macro
grid using the method:
\begin{lstlisting}
  void insertVertex ( const Dune::FieldVector<ctype, dimensionworld> &coord,
                      VertexId globalId );
\end{lstlisting}
This unique numbering is sufficient to use the grid factory concept in parallel.
Notice that elements and boundaries are inserted using a local vertex number
corresponding to the insertion order. \code{VertexId} in the current implementation is
an unsigned integer. 

To further increase efficiency, faces on process boundaries can also be inserted,
reducing the need for global communication during grid construction.
Similar to the \code{insertBoundarySegment} method, the grid factory in
\dune[ALUGrid] allows the insertion of process borders through the method
\begin{lstlisting}
  void insertProcessBorder ( const std::vector<unsigned int> &vertices );
\end{lstlisting}
While it is not necessary to insert process borders, we strongly recommend
doing so, because the construction of this information within the grid factory
requires an expensive global communication.
Note that this method will not work accurately, if it is called for some process
borders only.

In some cases it is easier to simply insert into the factory that a certain
face of an element is on the border or on the boundary (see the example in
Section~\ref{sec:StructuredGridFactory}).
The grid factory in \dune[ALUGrid] allows this through the following methods:
\begin{lstlisting}
  void insertBoundary ( int element, int faceInElement );
  void insertProcessBorder ( int element, int faceInElement );
\end{lstlisting}
The local face numbering used for \code{faceInElement} corresponds to the \dune
reference element.

\subsubsection{StructuredGridFactory}
\label{sec:StructuredGridFactory}

An example of how to use the new methods on the grid factory to construct a distributed grid 
is provided in the specialization of the \code{StructuredGridFactory} in
\file{dune/alugrid/common/structuredgridfactory.hh}.

Given an interval $[a,b]\subset\setR^3$ and a subdivision vector $N \in \setN^3$, a
distributed Cartesian grid is constructed.
Each process first uses \code{SGrid} (a structured grid manager available in
\dune[grid]) to setup a Cartesian grid.
A space filling curve is then used to partition this grid and the
distributed grid is constructed using the extended grid
factory of \dune[ALUGrid] on each process. 
Note that the resulting partition on each process does not consist of a product
of intervals, since the distribution is done using a space filling curve.

The following code snippet shows the idea in a very general setting.
The \code{gridView} object is the leaf grid view of a given grid (e.g. 
of a \code{SGrid}), \code{indexSet} denotes its index set, and the 
\code{partitioner} object provides a method \code{rank(const Entity &)}
returning the MPI rank that the entity shall be assigned to (e.g. based on a
space filling curve).
\begin{lstlisting}
// create ALUGrid GridFactory
GridFactory< Grid > factory;

// map global vertex ids to local ones
std::map< IndexType, unsigned int > vtxMap;

const int numVertices = (1 << dim);
std::vector< unsigned int > vertices( numVertices );

int nextElementIndex = 0;
const auto end = gridView.template end< 0 >();
for( auto it = gridView.template begin< 0 >(); it != end; ++it )
{
  const Entity &entity = *it;
  if( partitioner.rank( entity ) != myrank )
    continue;

  // insert vertices and element
  const typename Entity::Geometry geo = entity.geometry();
  for( int i = 0; i < numVertices; ++i )
  {
    const IndexType vtxId = indexSet.subIndex( entity, i, dim );
    auto result = vtxMap.insert( std::make_pair( vtxId, vtxMap.size() ) );
    if( result.second )
      factory.insertVertex( geo.corner( i ), vtxId );
    vertices[ i ] = result.first->second;
  }
  factory.insertElement( entity.type(), vertices );
  const int elementIndex = nextElementIndex++;

  const auto iend = gridView.iend( entity );
  for( auto iit = gridView.ibegin( entity ); iit != iend; ++iit )
  {
    const Intersection &isec = *iit;
    const int faceNumber = isec.indexInInside();
    // insert boundary face in case of domain boundary
    if( isec.boundary() )
      factory.insertBoundary( elementIndex, faceNumber );
    // insert process boundary if the neighboring element has a different rank
    if( isec.neighbor() && (partitioner.rank( *isec.outside() ) != myrank) )
      factory.insertProcessBorder( elementIndex, faceNumber );
  }
}
\end{lstlisting}

\subsubsection{Dune Grid Format (DGF)}
\label{sec:paralleldgf}

The \code{DGFParser} has also been extended to make use of the parallel
grid construction available in \dune[ALUGrid]. For each process a
\code{dgf} file (e.g., \file{grid.dgf.P.1}, ...,\file{grid.dgf.P.P}) is used
containing only one part of the grid. As in the serial case the blocks with the
information on the elements uses a process local numbering of the vertices. 
A new block \code{GlobalVertexIndex} has to be added, where a globally unique integer for
each vertex in this partition is provided in the same order used for the coordinates in
the \code{Vertex} block.
The file passed to the \code{GridPtr} class 
(e.g. \file{grid.dgf.P}) contains only the block \code{ALUParallel} listing the
file names of the individual partitions for each process.

The following shows an example for the domain $[0,1]^3$ divided into $4$ elements and distributed over
two processors:

\begin{center}
\begin{tabular}[t]{lll}
\file{cube.dgf.2} & \file{cube.dgf.2.1} & \file{cube.dgf.2.2} \\
\begin{lstlisting}[boxpos=t]
  DGF
  ALUPARALLEL
    cube.dgf.2.1
    cube.dgf.2.2
  #
\end{lstlisting} &
\begin{lstlisting}[boxpos=t]
DGF
VERTEX
0 0 0
0.5 0 0
0 0.5 0
0.5 0.5 0
0 0 1
0.5 0 1
0 0.5 1
0.5 0.5 1
0 1 0
0.5 1 0
0 1 1
0.5 1 1
#
CUBE
0 1 2 3 4 5 6 7
2 3 8 9 6 7 10 11
#
GLOBALVERTEXINDEX
0
1
2
3
4
5
6
7
12
13
14
15
#
\end{lstlisting} &
\begin{lstlisting}[boxpos=t]
DGF
VERTEX
0.5 0 0
1 0 0
0.5 0.5 0
1 0.5 0
0.5 0 1
1 0 1
0.5 0.5 1
1 0.5 1
0.5 1 0
1 1 0
0.5 1 1
1 1 1
#
CUBE
0 1 2 3 4 5 6 7
2 3 8 9 6 7 10 11
#
GLOBALVERTEXINDEX
1
8
3
9
5
10
7
11
13
16
15
17
#
\end{lstlisting}
\end{tabular} 
\end{center}

These files were generated by using the utility
\code{ParallelDGFWritter} class (in
\file{dune/alugrid/common/writeparalleldgf.hh}) to provide distributed dgf
files from a given input dgf file. 

\subsubsection{Utility programs}

The \dune[ALUGrid] module also provides an utility program \file{utils/convert-macrogrid/convert} 
to convert a normal DGF file or a legacy \alugrid
macro grid file into \dune[ALUGrid]'s new binary or compressed binary macro grid file format. 
This tool can also decompose the macro grid into several partitions. 
\dune[ALUGrid] is able
to read decomposed macro grids if the number of partitions of the macro grid is smaller
or equal to the used number of cores. This is especially useful for very
large macro grids which will not fit into the memory of a single core. In addition the compressed
binary format reduces storage requirements and decreases storage access times.

\subsection{Backup and Restore}
\label{sec:dataio}

For backup and restore as it is needed for checkpointing and postprocessing 
a new interface was recently introduced into \dune[Grid]. To our knowledge
\dune[ALUGrid] is the first grid manager implementing this interface so we will
go into a bit more detail in the following.
The interface is given by
\begin{lstlisting}
  template< int dim, int dimworld, 
            ALUGridElementType elType, 
            ALUGridRefinementType refineType, class Comm >
  struct BackupRestoreFacility< 
            ALUGrid< dim, dimworld, elType, refineType, Comm > >
  {
    /** perform backup of grid to given std::ostream */
    static void backup ( const Grid &grid, std::ostream &stream ) ;

    /** restore grid from std::istream and return pointer to
        newly created grid object */
    static Grid* restore ( std::istream &stream ) ;
  };
\end{lstlisting}

The \code{BackupRestoreFacility} provides two further \code{backup} and \code{restore} methods 
where a filename is the argument instead of a stream. These methods have been added for 
legacy codes like \alberta \cite{alberta:05} that might not support the read and write via streams. 
For \dune[ALUGrid] these are simply implemented using a file stream and the calling the
above mentioned methods.

For data I/O on large parallel machines we provide two mechanisms. The conventional
approach is to use standard file streams to create a binary file for each process containing the macro
grid cells, refinement tree, and index information for the corresponding partition. 
This becomes very cumbersome when the code is used with many cores. 
Therefore, the second approach is to use a \code{std::stringstream} to write all information into
a buffer of type \code{char*} and then use a library like \sionlib \cite{sionlib} to
write the data to the storage unit. This approach has the advantage that libraries like
\sionlib provide the maximal I/O performance but do not limit 
\dune[ALUGrid] to be used only with this library. For libraries that require the size
of data to be written, like \sionlib, the intermediate storage in a
\code{char} buffer is necessary since for the adaptive grid the number elements is not
known apriori. How \sionlib is used is shown in the examples presented in 
\file{examples/backuprestore}. This example explains how backup/restore is done
using different ways to write data to the storage device.

Note that \dune[ALUGrid] will only backup/restore it's \code{LocalIdSet}. The
\code{GlobalIdSet} is generated from the unique macro element id (built from the unique vertex ids) 
and the position in the refinement tree and therefore does not need to be stored explicitly.
Furthermore, the persistent order of the macro grid automatically induces the
same traversal order for the hierarchical grid.
Since both, the \code{LevelIndexSet} and the \code{LeafIndexSet} are 
generated by grid traversal and \textit{insert on first visit} strategy, both index set variants 
preserve their indices over a backup and restore process.

\subsection{Overlapping Communication and Computation}
\label{sec:communication}
In a numerical algorithm, degrees of freedom are typically attached to grid
entities.
Now, a single grid entity can be visible to multiple processes and any data
attached to it needs to be synchronized between these processes.
The \dune\ grid interface therefore requires each grid view to support this
synchronization through a \code{communicate} method:
\begin{lstlisting}
  template< class DataHandle >
  void communicate ( DataHandle &, InterfaceType,
                     CommunicationDirection ) const;
\end{lstlisting}
The interface type and communication direction specify the set of entities on
which data has to be sent or received.
On the sending side the data handle is responsible for packing entity data into
a buffer; on the receiving side it unpacks the data again.
The actual data transfer is done transparently by the grid implementation.

After all data has been sent, the grid implementation has to wait until incoming
data is received, which can be a waste of valuable computation time.
Indeed, many numerical algorithms can be split into work that depends on the
shared data and work that does not.
The latter part can actually be done while communication is in progress
simply by splitting sending and receiving in two parts and is supported even by
the oldest MPI implementations.

To make use of the valuable communication time, \dune[ALUGrid] allows to delay
the receiving process to a convenient point in the algorithm.
The actual communication initiated by \code{communicate} becomes an object:
\begin{lstlisting}
  template< class DataHandle >
  Communication< DataHandle > communicate ( DataHandle &, InterfaceType,
                                            CommunicationDirection ) const;
\end{lstlisting}
Such a \code{Communication} object satisfies the following interface:
\begin{lstlisting}
  struct Communication
  {
    // wait for communication to finish if not already done
    ~Communication () { if( pending() ) wait(); }
    
    // is this communication still pending?
    bool pending () const;

    // wait for communication to finish
    void wait ();
  };
\end{lstlisting}
While the communication is pending, i.e., while wait has not been called, the
reference to the data handle must remain valid.
As \code{wait} is automatically called in the destructor, ignoring the return
value will result in a blocking communication. Thus no change is required
to existing code if blocking communucation is to be used.

If \code{gridView} is a grid view of an \code{ALUGrid} object, overlapping communication
and computation is rather simple:
\begin{lstlisting}
  // construct data handle for the communication
  auto comm = gridView.impl().communicate ( dataHandle, interface, dir );
  // do some computation not depending on the remote data
  comm.wait();
  // do computation depending on the remote data
\end{lstlisting}
Note that the method \code{impl} is only available if
experimental grid extensions have been enabled in \dune[Grid]
and would no longer be required once the new interface is added into \dune[Grid].

A possible usage of the communication hiding 
is presented in the following code snippet.
The method is implemented in \file{examples/communication} where the main
change in the time loop is quite simple:
\begin{lstlisting}
  // original non-blocking code: dt = scheme( time, solution, update ) ;
  {
    // new code: compute data on border and ghost entities
    dt = scheme.border( time, solution, update );
    // start non-blocking communication
    auto commObject = grid.communicate( handle, interface, direction );
    // do computation not depending on remote data 
    dt = std::min(dt , scheme( time, solution, update ) );
  } // communication will be finished when commObject goes out of scope
\end{lstlisting}

\subsection{Adaptation Using Call-Backs}
\label{sec:adaptcallback}
Grid modification in \dune is performed in three steps. First
\code{grid.preadapt()} is called to start the modification phase. After
this method has been called the index sets are no longer valid
and data has to be accessed based either on one of the \code{IdSets} or using a
\code{PersistentContainer}. Both allow storage of data persistently during
grid modification and on the whole hierarchy of the grid making it possible for data
to be restricted and prolongated from one level to another. Next 
\code{grid.adapt()} is called which refines or coarsens grid elements according 
to markers set by the user. Finally \code{grid.postadapt()} is called,
ending the modification phase and reinitializing the \dune consecutive,
zero starting index sets allowing to store user data in consecutive memory
locations.

The main steps for the user consist in making data persistent
during the modification stage of the grid, prolongation of data if
elements are refined, and restriction of data if elements are coarsened. 
A common approach is to store the data in a vector-like structure in the
computation phase for efficient memory access.
The necessary copying of the data into a \code{PersistentContainer} during
the modification phase makes this step computationally more expensive.
Alternatively, the user can store data directly in a \code{PersistentContainer} 
which means that the storage does not have to be modified during grid changes but
sacrificing efficiency during the computation phase due to more expensive data
access.
In \dune the \code{PersistentContainer} can be specialized
for each grid implementation. A default implementation uses a
\code{std::map} to store the data using the \code{LocalIdSet} of the grid
as key. \dune[ALUGrid] uses a speciallization of this class based on a
\code{std::vector} to store the data. Each entity stores an integer which
is unique within the grid hierarchy and which can be used to access the
data within the vector. In contrast to a \dune \code{IndexSet} this index
is not necessarily zero starting and consecutive, resulting in holes within
the \code{PersistentContainer} but allowing for a constant retrieval time
of the data. In our example adaptive Finite Volume scheme the two storage strategies
are available for testing. By default the \code{PersistentContainer} is used
but by defining \code{USE_VECTOR_FOR_PWF} the degrees of freedom will be
stored in a vector-like structure and moved into a
\code{PersistentContainer} only during the grid modification stage.
In all our tests the storage of data in the \code{PersistentContainer} was
significantly more efficient. 
Some results are shown in the \dune columns of Table~\ref{tab:checkadaptation}.

In addition to the approach described above, \dune[ALUGrid] provides an
adaptation mechanism using a callback approach, similar to the \dune
communication and loadbalancing interface
(see Section \ref{sec:userdeflb}).
Instead of the \code{grid.preAdapt()}, \code{grid.adapt()}, \code{grid.postAdapt()}
algorithm, a single call to \code{grid.adapt( dataHandle )} is required. 
The \code{dataHandle} has to be derived from 
\begin{lstlisting}
  template< class Grid, class Impl >
  struct AdaptDataHandle
  {
    typedef typename Grid::template Codim< 0 >::Entity Element;

    void preCoarsening  ( const Element &father );
    void postRefinement ( const Element &father );
  };
\end{lstlisting}

The method \code{preCoarsening} is called on the element \code{father} before all its
descendants are removed. Accordingly, the method \code{postRefinement} is called
immediately after descendants for an entity \code{father} are created.
Since these methods are called during grid modification the
\code{IndexSets} on the grid are not available and data has to be stored in
some peristent manner, e.g., using the \code{PersistentContainer}. There is
no need to call \code{preAdapt(),postAdapt()} on the grid.

This variant of the adaptation cycle is implemented in 
\file{examples/callback/adaptation.hh}.
Assuming that the degrees of freedom are stored in a
\code{PersistentContainer} one simply needs to call
\begin{lstlisting}
grid_.adapt( *this );
\end{lstlisting}
and implement the two callback methods
\begin{lstlisting}
  void preCoarsening ( const Entity &father )
  {
    Container &container_ = getSolution().container();
    // average the data from all children and copy onto the father entity
    Vector::restrictLocal( father, container_ );
  }

  // called when children of father where newly created
  void postRefinement ( const Entity &father )
  {
    Container &container_ = getSolution().container();
    container_.resize();
    // copy the data from the father onto all its children
    Vector::prolongLocal( father, container_ );
  }
\end{lstlisting}

The results of using the callback approach are shown in the corresponding
columns of Table~\ref{tab:checkadaptation}.
In summary, our tests indicate a gain of up to $10\%$ using callback
adaptation. 
In addition the overall implementation is simpler since the
hierarchic restriction and prolong methods do not have to be implemented. 
To run the test described here go to the \file{examples/callback} directory and run the 
\code{check-adaptation.sh} script. The implementation with a
\code{PersistentContainer} is also compared here with the version based on
vector-like structure. The advantage of using a
\code{PersistentContainer} for the degrees of freedom in the Finite Volume
scheme is significant (more than $20\%$).

\begin{table}[ht]{}
\renewcommand{\arraystretch}{1.5}
\begin{center}
\caption{Results for callback adaptation and dof storage strategy obtained on 
a single core from our small cluster.
See script \file{examples/callback/check-adaptation.sh}. 
\textbf{T} stands for transport problem and \textbf{E} for Euler problem, followed 
by the three program parameters used.}
\begin{tabular}{rr|cccc}
\multicolumn{2}{r|}{storage} & \code{vector} & \code{vector} & \code{PersistentContainer} & \code{PersistentContainer} \\ \hline
\multicolumn{2}{r|}{adaptation} & \dune & callback & \dune & callback \\ 
\hline\hline
 \textbf{T} & \code{2 0 2} & 251s & 227s & 194s & 173s \\ \hline
 \textbf{T} & \code{2 0 3} & 2411s & 1820s & 2222s & 1647s \\ \hline
 \textbf{E} & \code{21 0 3} & 106s & 83s & 99s & 77s     \\ \hline
 \textbf{E} & \code{21 0 4} & 1070s & 1037s & 833s & 766s \\
\end{tabular}
\label{tab:checkadaptation}
\end{center}
\end{table}

\subsection{Internal Load Balancing}
\label{sec:userdeflb}
There are two stages in a computation where load balancing is essential in a simulation.
During the start up phase of the computation where the grid has to be distributed over the
available number of processes and after the grid has been locally refined. Even if the
grid has been partitioned beforehand and \dune[ALUGrid]'s parallel grid factory is used,
it is still sometimes of practical interest to repartition the grid after creation,
e.g., if a larger number of processes are available for the computation.
To this end the \dune[Grid] interface provides the method
\begin{lstlisting}
  bool loadBalance();
\end{lstlisting}
Even if the initial grid is optimally distributed, the load can become
unbalanced during the computation for example if local adaptivity is used.
In this case the method mentioned above is not sufficient as it does not allow to
migrate user data together with elements from one process to another.
To manage data migration the \dune[Grid] interface provides a second method
\begin{lstlisting}
  template< class DataHandleImpl, class Data >
  bool loadBalance( CommDataHandleIF< DataHandleImpl, Data > &dataHandle );
\end{lstlisting}
The handling of user data is achieved by a callback mechanism using the same 
interface used for communication during the computation. Basically, for each
element to be removed on the given process a method \code{gather} is called (to collect data to be
shipped with the element) and when a new element is added to the grid 
on the process then a method 
\code{scatter} is called (to deliver the data that was shipped with the element ) 
on the \code{dataHandle} instance. 

The main problem with these two methods is that there is no mechanism for
the user to intervene with the details of partitioning computed by the grid
manager. This new module provides two mechanisms for the user to improve the
internal load balancing to suit the need of the application at hand.
Before presenting these improvements, we give a brief description of how
\dune[ALUGrid]'s internal load balancing strategy works.

\dune[ALUGrid] only allows for horizontal load balancing,
i.e., partitioning of the elements on the macro level, migrating the whole 
tree below a given macro element from one process to another. Each macro
element $E$ is assigned a weight equal to the number of leaf elements below
$E$. Using these weights either a space filling curve approach is used 
or a graph partitioning algorithm is used. 
In \alugrid 1.52 only serial graph partitioning using the \metis library
\cite{metis} could be used. The serial graph partitioning requires the communication 
of the whole assembled graph to all processes which does not scale in terms of memory
and communication time.
While this method can still be used in \dune[ALUGrid], additional bindings to
\zoltan \cite{zoltan} have been added (providing space filling curve and
graph partitioning methods). 
Via the \zoltan interface \parmetis \cite{parmetis} is available as well.
The graph is constructed using the weighted macro elements as nodes and connecting 
neighboring macro elements $E_1,E_2$ with an edge in the graph. These edges are assigned 
weights according to the number of leafs below $E_1,E_2$ which are
neighbors. The node weights are to represent the computational cost, while
the edge weights represent the communication size in the case that these
elements come to be on different processors. 
The newly implemented partition algorithm for 
space filling curves makes \dune[ALUGrid] more self contained. 
As a default we are using the Hilbert space filling curve (see for example \cite{bader:13}) provided by \zoltan \cite{zoltan}.
The element weights described above are used to determine the optimal
partitioning of the space filling curve. Besides the space filling curve based 
approach provided by \zoltan called HSFC (id 13) \dune[ALUGrid] also provides it's own load
distribution algorithm. In this case it is assumed that the elements of the macro mesh
are sorted by a space filling curve. Then the distribution of the load boils down to the 
distribution of a 1d graph with attached weights. If \zoltan is available
and no pre-ordered mesh is provided, the Hilbert space filling curve from the \zoltan
package is used to sort the elements. As a fallback \dune[ALUGrid] also provides it's
on implementation based on the Z-curve (aka Morton curve) approach. 
The algorithm to partition the 1d graph is based on the one described in
\cite[Algorithm 16]{burstedde:11} with some slight modifications such as avoiding empty partitions in any
case if the number of macro elements is larger than the number of cores used.
\dune[ALUGrid]'s internal space filling curve with linkage (id 4) algorithm comes with a further advantage: 
The communication after a redistribution to identify master-slave node relations can be
done without communication. In all other cases listed in Table \ref{tab:lbmethods} an all-to-all communication
is needed to compute the master-slave relation of vertices that are present on multiple cores.

The simplest way to tweak the internal load balancing algorithm are three
parameters read from a file called \file{alugrid.cfg}, which
is searched for in the current working directory.
This file has to contain three values. 
The fist two numbers (\code{lbUnder}, \code{lbOver}) 
in the \file{alugrid.cfg} file allow to specify a
certain amount of load inbalance which has to be exceeded before the
partitioning is adjusted. A new partitioning is computed only if the
maximum number of leaf elements in a partition exceeds 
\code{lbOver} times the mean number of elements or the minumum number is
smaller than \code{lbUnder} times the mean number of elements in all
partitions.
The third value is an integer between
$0$ and $15$ determining the partitioning method to use.
Table~\ref{tab:lbmethods} gives an overview of available methods and their
numbering. 

\begin{table}
\renewcommand{\arraystretch}{1.25}
\caption{Internal partitioning methods and corresponding id. 
}
\label{tab:lbmethods}
\begin{center}
\begin{tabular}{lc|lc}
  method name & id  & method name & id \\ \hline \hline  
NONE & 0 & COLLECT (to rank $0$) & 1 \\
  Space Filling Curve (linkage) & 4  
  &Space Filling Curve & 9 \\
METIS (PartGraphKway) & 11
  & METIS (PartGraphRecursive) & 12 \\
ZOLTAN (HSFC) & 13 
  & ZOLTAN (GRAPH) & 14 \\
ZOLTAN (PARMETIS) & 15
\end{tabular}
\end{center}
\end{table}

A second option to influence the outcome of the load balancing algorithm is
to provide other weights for the elements (i.e., the graph nodes).
This can improve the overall
effciency of a scheme if the number of leaves does not directly represent
the computational cost associated with a given macro element. An example
for this are reactive flow problems where substepping in time is used to
resolve stiff sources locally on each element \cite{gessner:01}.
Further examples are the solution of PDEs in a moving domain \cite{motor:13} or 
a multi-domain approach where partial differential equations with different
complexity are solved in different domains represented on the same underlying grid~\cite{multidomain:12}. 
The corresponding additional methods are
\begin{lstlisting}
  template< class LBWeights >
  bool loadBalance ( LBWeights &weights );
  template< class LBWeights, class DataHandleImpl, class Data >
  bool loadBalance ( LBWeights &weights, 
                     CommDataHandleIF< DataHandleImpl, Data > &dataHandle );
\end{lstlisting}
\code{LBWeights} must implement \code{int operator()(const Grid::Codim<0>::Entity &)}
which will be called for each macro element to provide the weight, here an integer value.
An example usage is shown in
\file{examples/loadbalancing/loadbalance_simple.hh}. Each leaf element is assumed to 
carry a computational cost of $2^l$ where $l$ is the level of the leaf element.
The weight for a macro element is then simply the sum of the weights over all underlying leaf elements.

\begin{figure}[!ht]
  \subfloat[overall]{\label{fig:scaling314_simp}
  \includegraphics[width=0.32\linewidth]{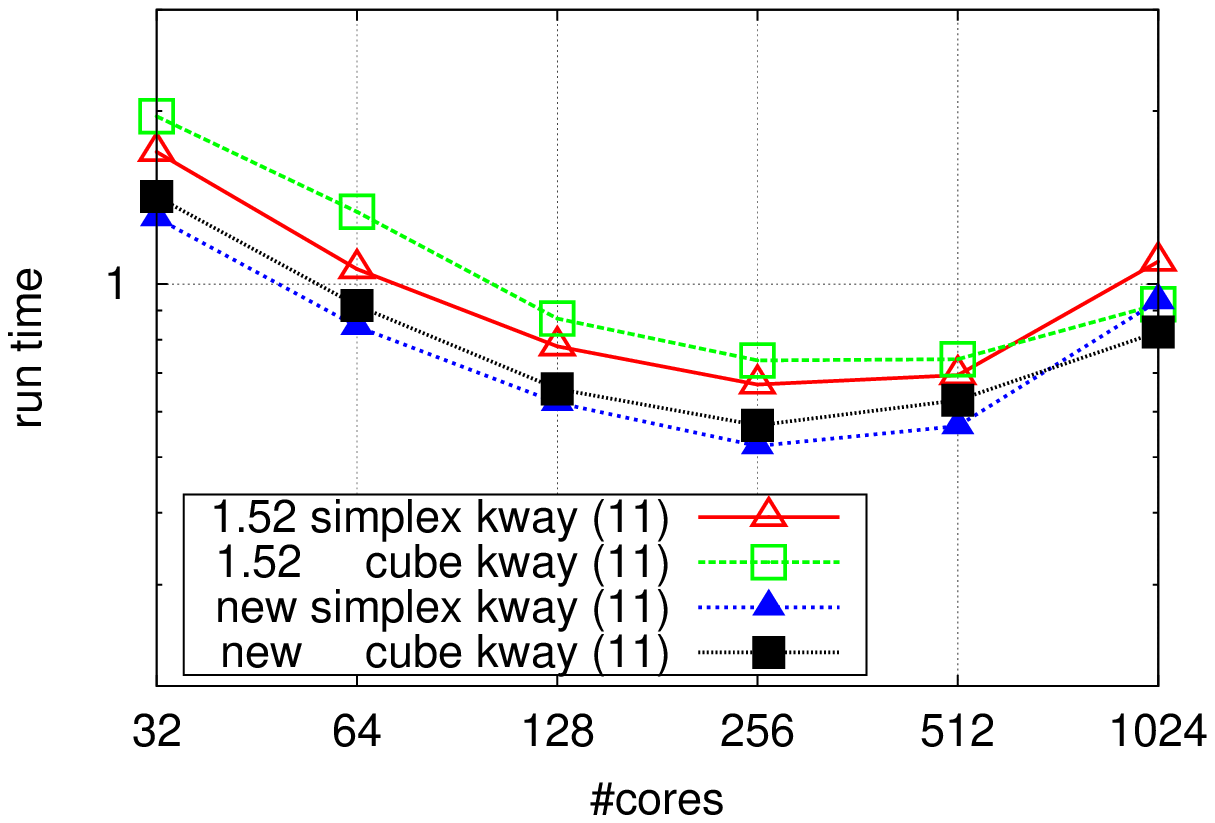}}
  \subfloat[adaptation]{\label{fig:scaling314_simp_ad}
  \includegraphics[width=0.32\linewidth]{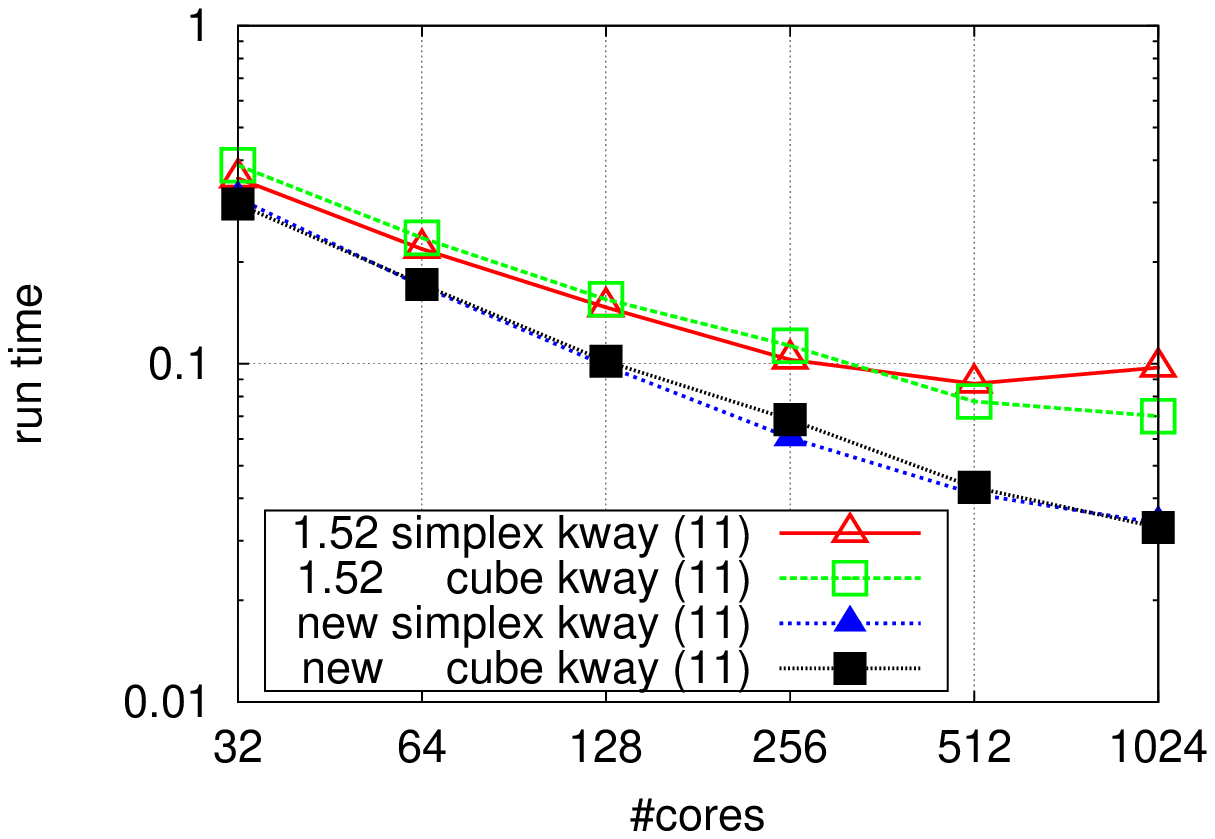}}
  \subfloat[load balancing]{\label{fig:scaling314_simp_lb}
  \includegraphics[width=0.32\linewidth]{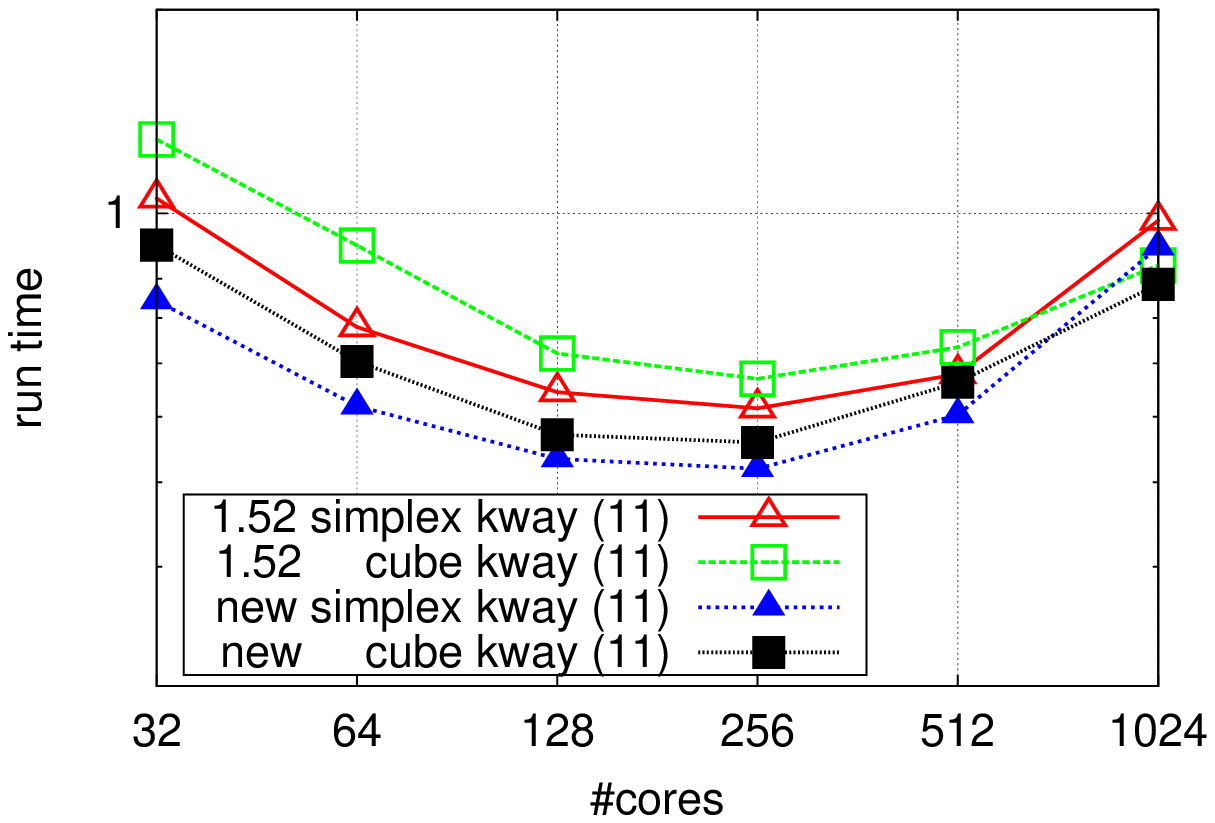}}
\caption{Comparison of \alugrid $1.52$ and \dune[ALUGrid] using the
         ball example with a macro mesh of $32\,768$ hexahedrons or $196\,608$ tetrahedrons.
         The grid is refined uniformly once and the maximal refinement level is
         $4$ (parameter 3 1 4  for example \code{main_ball}).
         Here, we only use the METIS PartGraphKway (partition method id 11)
         method for domain decomposition.}
\label{fig:scaling314_comparison}
\end{figure}

\begin{figure}[!ht]
 \subfloat[overall]{\label{fig:scaling314_simpconf}
  \includegraphics[width=0.32\linewidth]{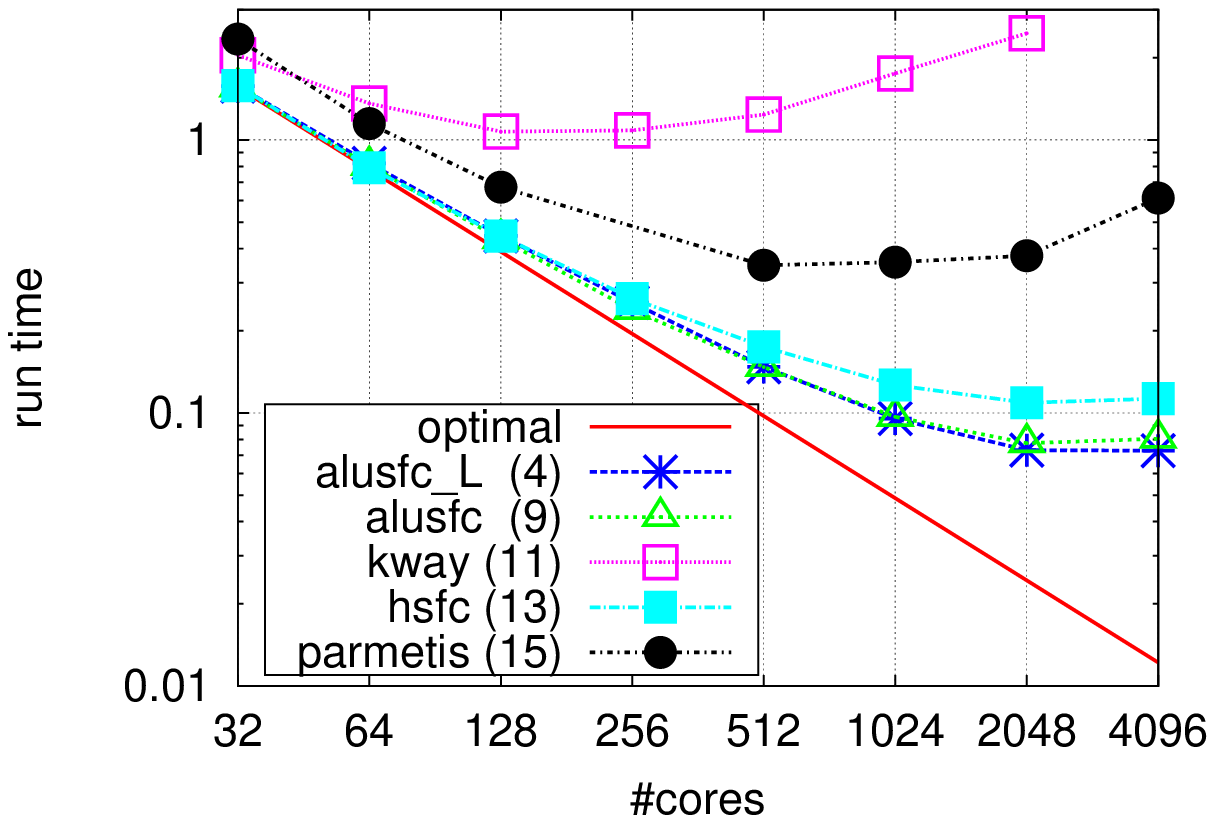}}
  \subfloat[adaptation]{\label{fig:scaling314_ad}
  \includegraphics[width=0.32\linewidth]{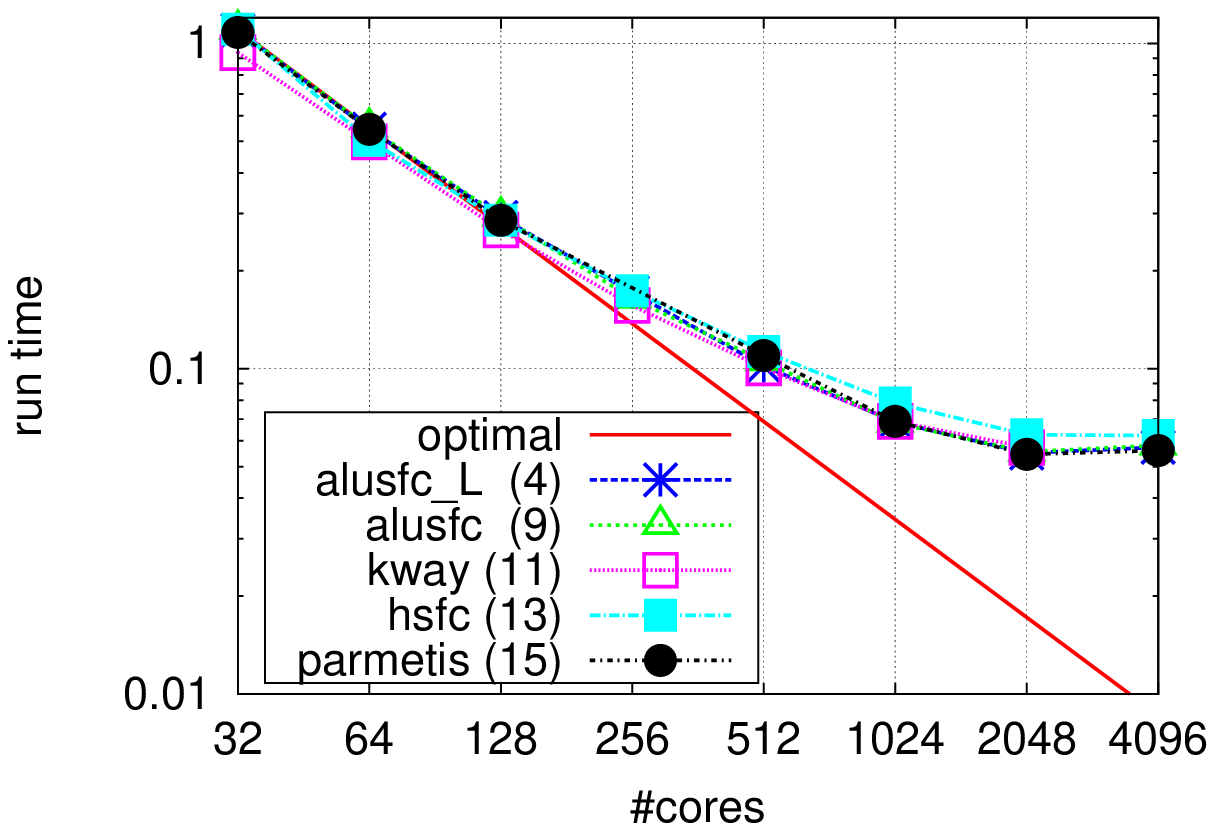}}
  \subfloat[load balancing]{
  \label{fig:scaling314_lb}
  \includegraphics[width=0.32\linewidth]{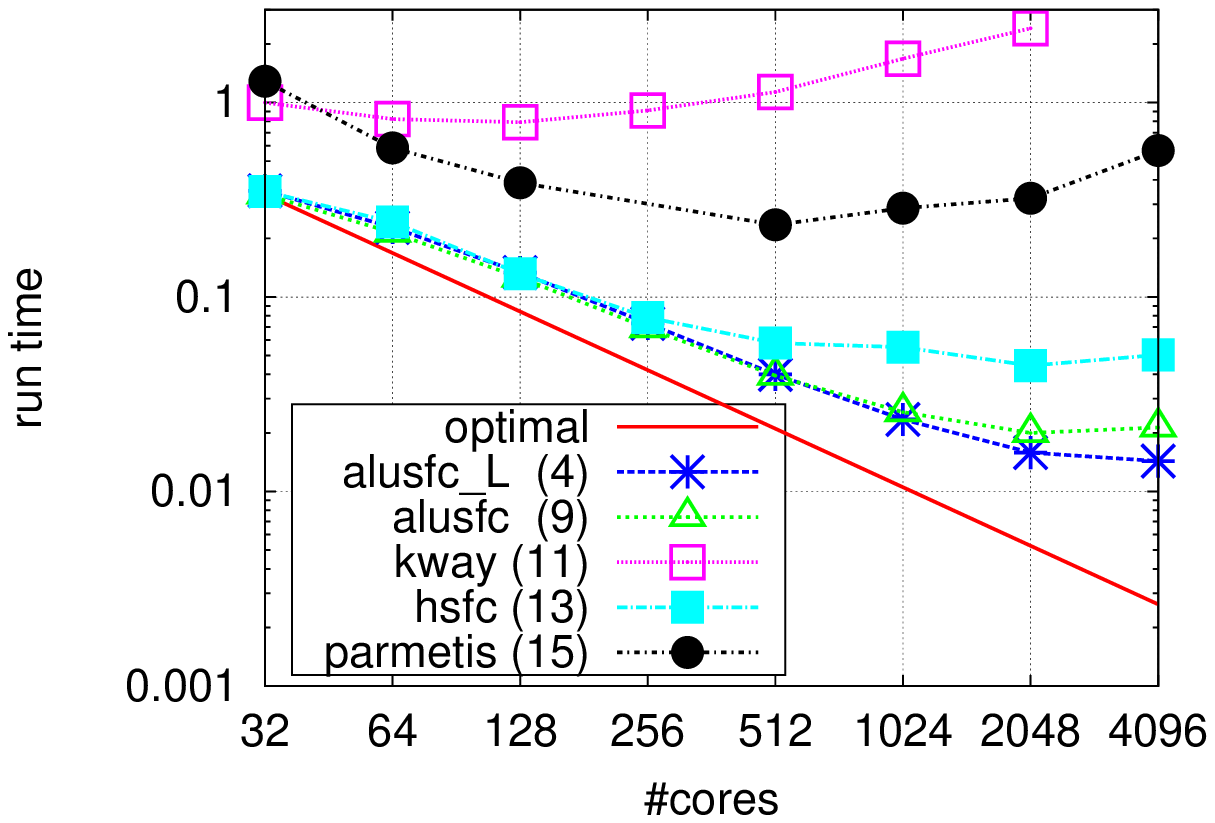}}
\caption{Strong scaling of the 
          ball example from equation \eqref{test:ball} using a conforming simplex grid with 
         a macro mesh containing $196\,608$ tetrahedrons.
         The grid is refined uniformly once and the maximal allowed refinement level is
         $4$ (parameter 3 1 4  for example \code{main_ball}).
         The graphs show the average run time per time step of different parts of the
         algorithm.}
\label{fig:scaling314_conf}         
\end{figure}

\begin{figure}[!ht]
  \subfloat[overall]{\label{fig:scaling314_cube_ov}
  \includegraphics[width=0.32\linewidth]{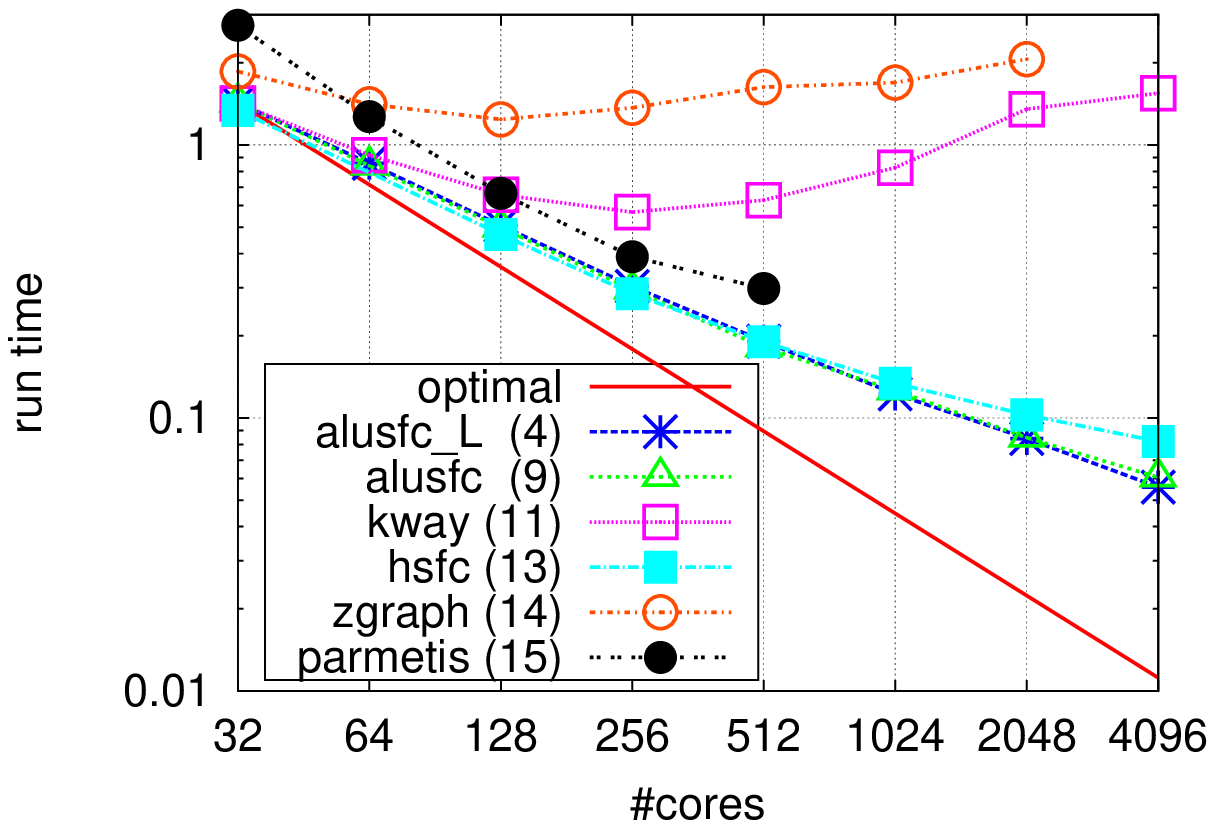}}
  \subfloat[adaptation]{\label{fig:scaling314_cube_ad}
  \includegraphics[width=0.32\linewidth]{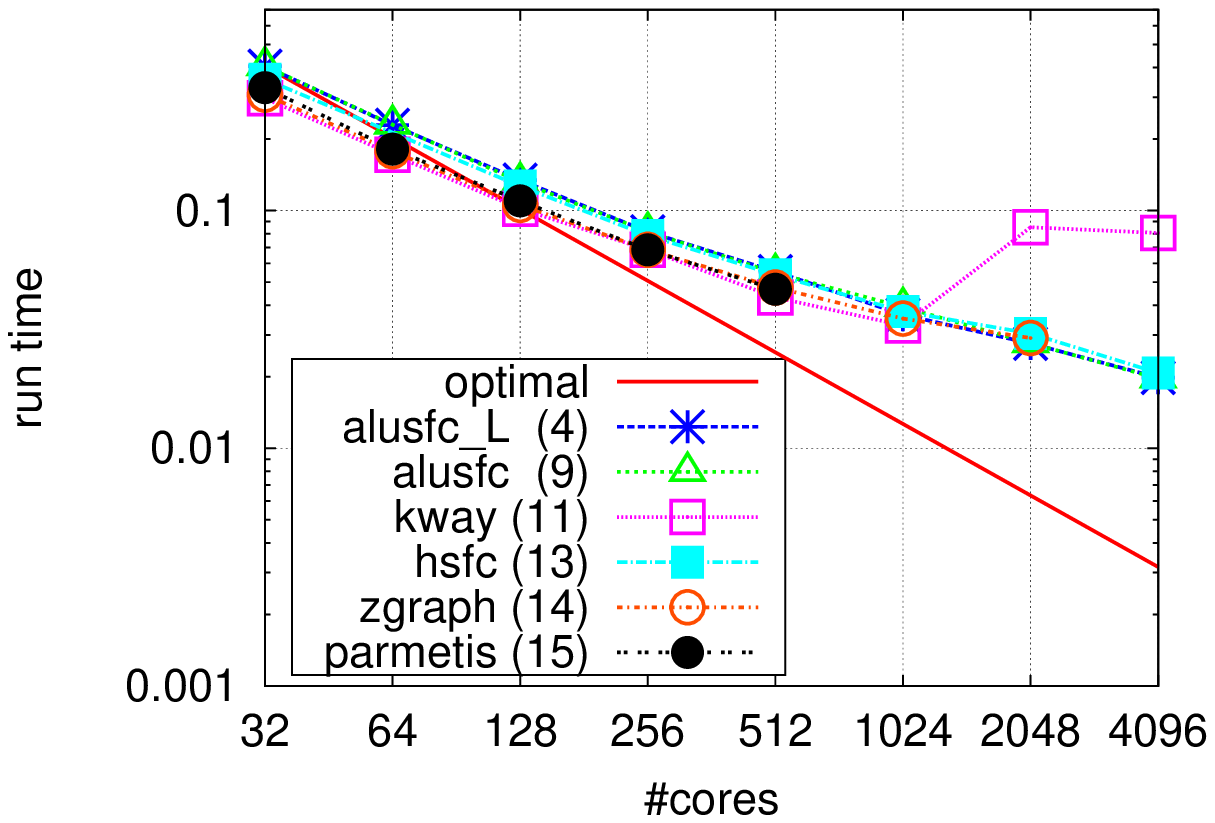}}
  \subfloat[load balancing]{\label{fig:scaling314_cube_lb}
  \includegraphics[width=0.32\linewidth]{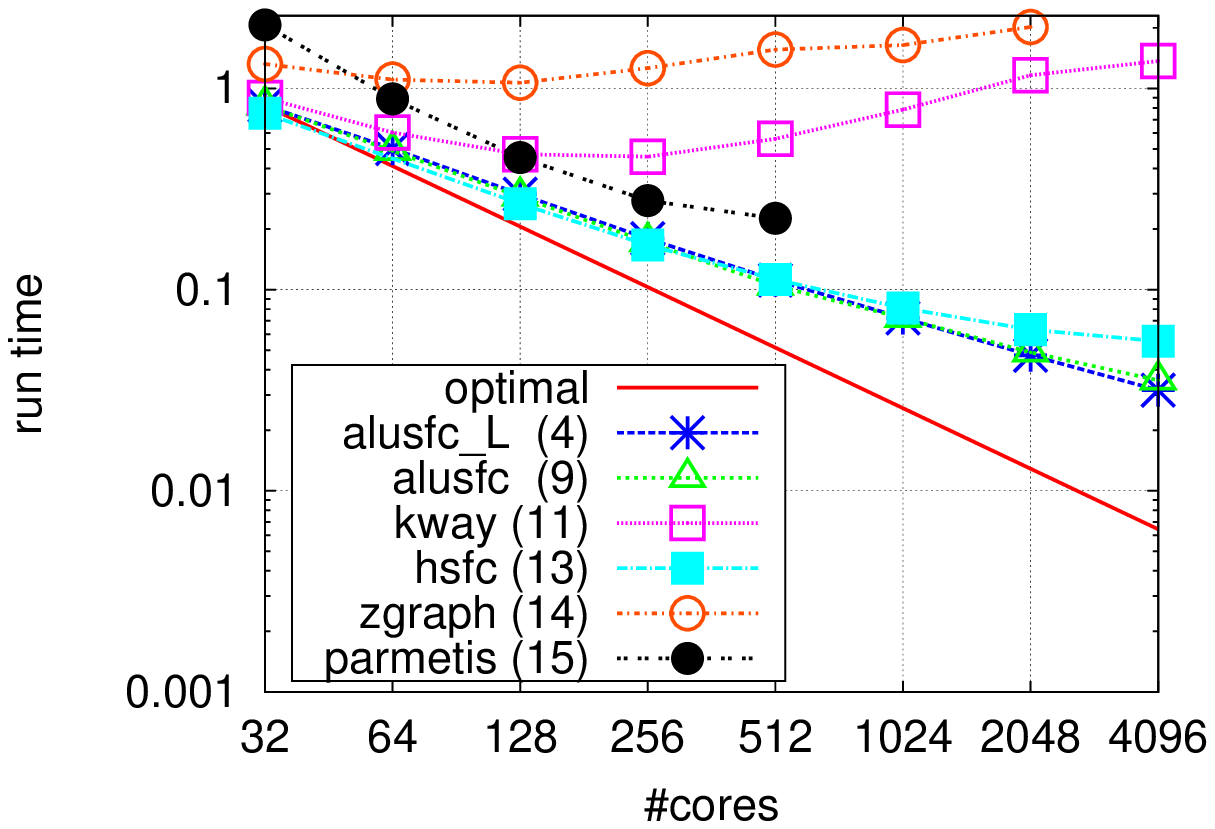}}
  \caption{Strong scaling of the ball example from equation \eqref{test:ball} using a non-conforming cube grid with
         a macro mesh containing $32\,768$ hexahedrons.
         The grid is refined uniformly once and the maximal allowed refinement level is
         $4$ (parameter 3 1 4  for example \code{main_ball}).
         The graphs show the average run time per time step of different parts of the
         algorithm.}
\label{fig:scaling314_cube}         
\end{figure}

\begin{figure}[!ht]
  \subfloat[overall]{\label{fig:scaling403_cube_ov}
  \includegraphics[width=0.32\linewidth]{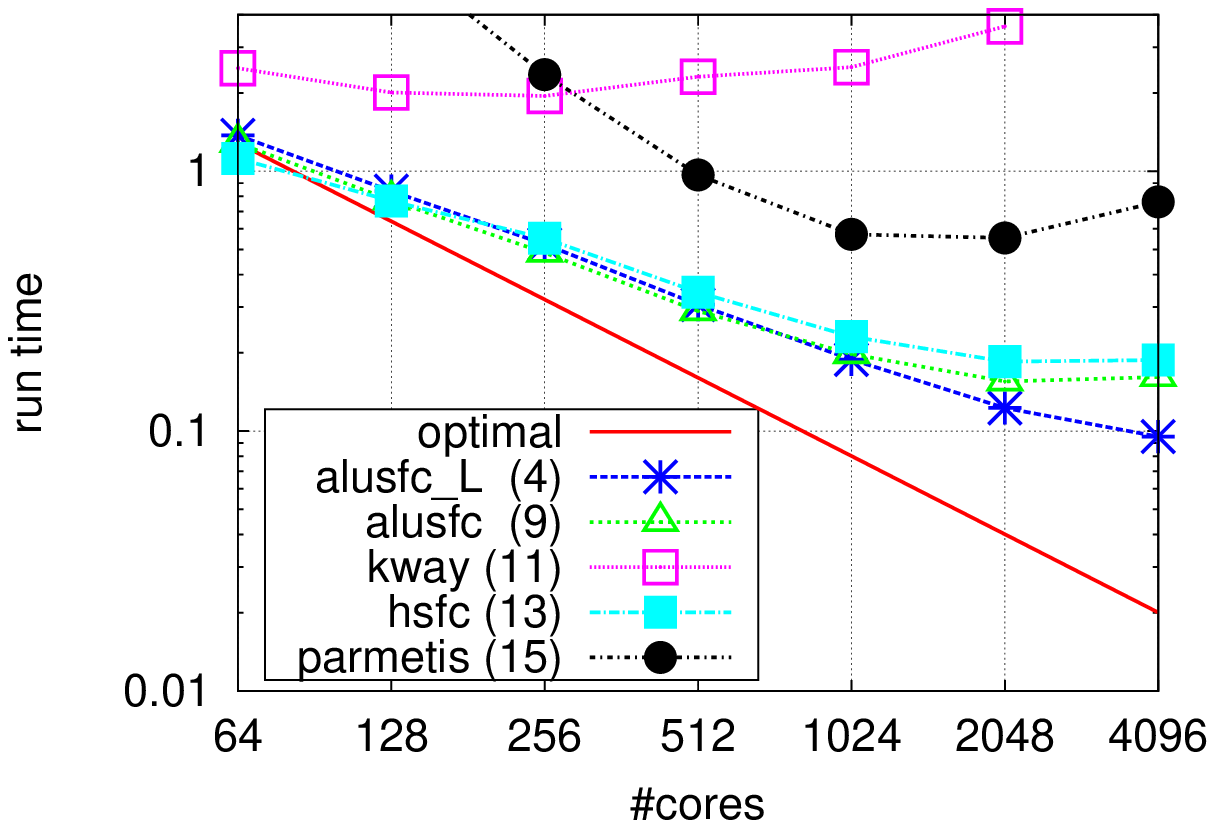}}
  \subfloat[adaptation]{\label{fig:scaling403_cube_ad}
  \includegraphics[width=0.32\linewidth]{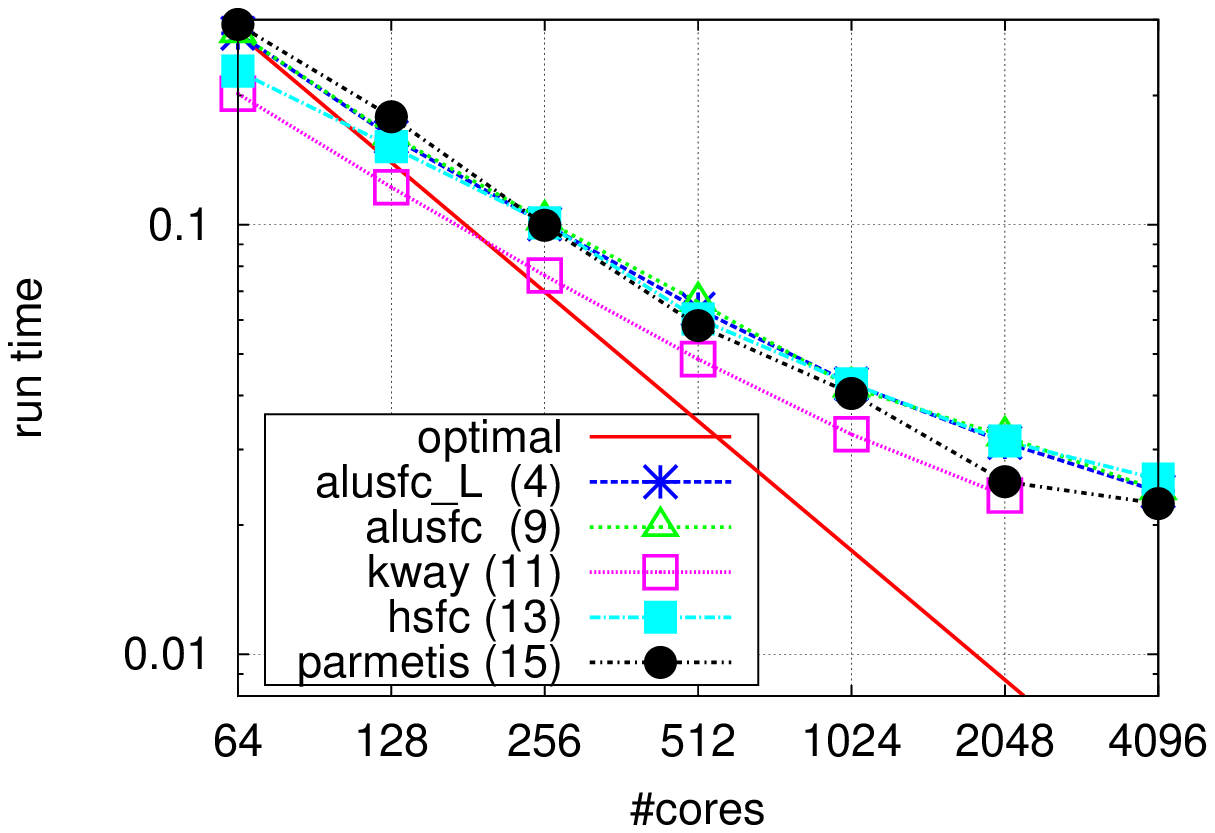}}
  \subfloat[load balancing]{\label{fig:scaling403_cube_lb}
  \includegraphics[width=0.32\linewidth]{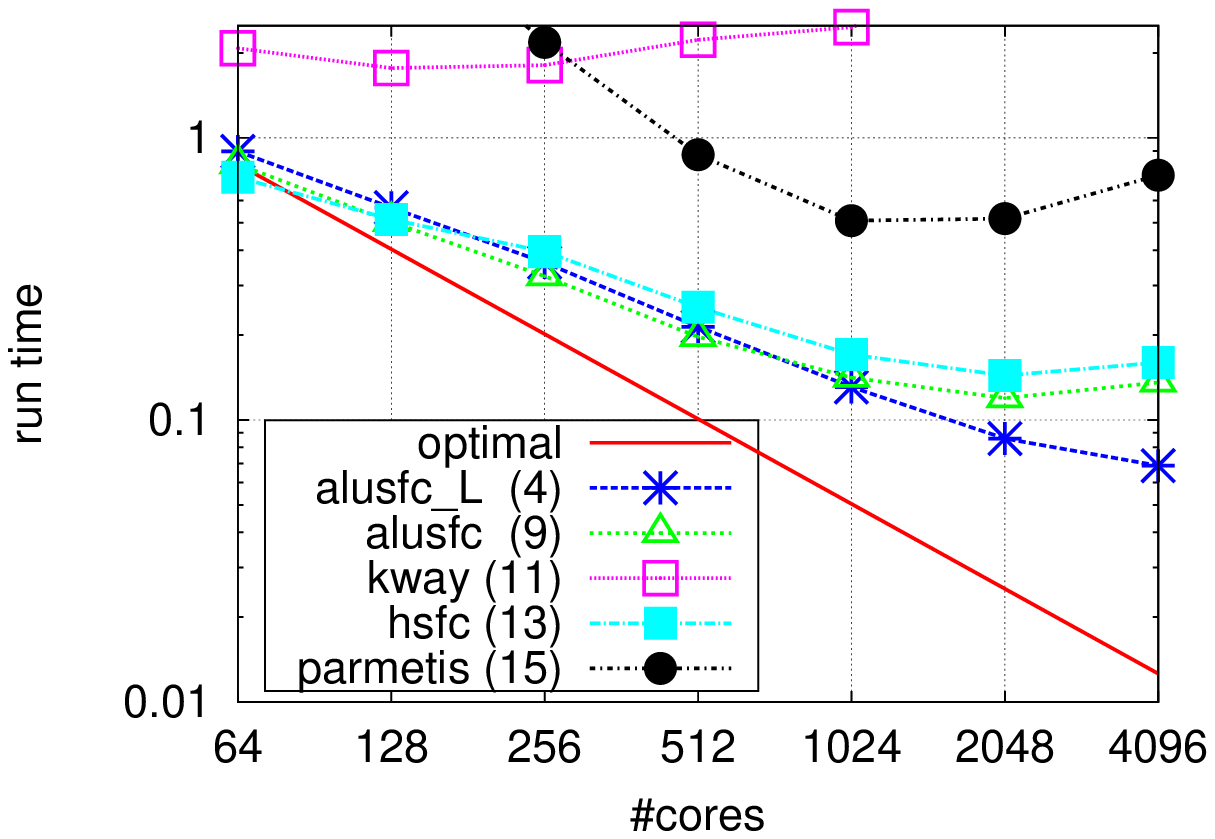}}
  \caption{Strong scaling of the ball example from equation \eqref{test:ball} using a non-conforming cube grid with
         a macro mesh containing $262\,144$ hexahedrons.
         The maximal allowed refinement level is
         $3$ (parameter 4 0 3  for example \code{main_ball}).
         The graphs show the average run time per time step of different parts of the
         algorithm.}
\label{fig:scaling403_cube}
\end{figure}

In Figure \ref{fig:scaling314_comparison}, \ref{fig:scaling314_conf}, \ref{fig:scaling314_cube}, and \ref{fig:scaling403_cube} 
we present a comparison of the different load balancing algorithms available in \dune[ALUGrid]. 
The results show a strong scaling study using the ball example with 
refinement as described in \eqref{test:ball}. 
The scaling studies have been carried out on Yellowstone \cite{Yellowstone}.

In Figure \ref{fig:scaling314_comparison} we present a comparison of run times for
\alugrid's $1.52$ version and the new \dune[ALUGrid] module. 
Since in version $1.52$ only \metis was available
for partitioning we only compare the run times using the \metis partitioning. 
We discover that both the tetrahedral and the hexahedral version perform better in the new
\dune[ALUGrid] implementation.

For the comparison of load balancing methods in 
Figure~\ref{fig:scaling314_conf}, \ref{fig:scaling314_cube}, and \ref{fig:scaling403_cube} 
we can see that the
space filling curve approaches, either \dune[ALUGrid]'s internal methods 
or the HSFC method from \zoltan, 
perform best even if the macro grid does not allow a good partitioning
anymore because the average element per core ratio is very small. The graph
partitioning methods are in general more expensive even though the created partitions 
seem to be more efficient in terms of communications effort resulting in faster run
times for the adaptation step (see Figure \ref{fig:scaling314_ad},
\ref{fig:scaling314_cube_ad}, and \ref{fig:scaling403_cube_ad}).
As a drawback all tested graph partitioning methods fail when the number of elements per core 
becomes very small. We have to point out that this example is 
heavily communication based and especially
the load balancing step, which is done in every time step, is very communication
intensive. So it seems even more impressive that the run times still drop when 
using $2048$ or $4096$ cores. 
This is confirmed by a comparable study in \cite{amdis2} where a stagnation in strong
scaling was observed when adaptivity and load balancing was done every time step. 
As a conclusion the space filling curve approaches seem 
more suitable for problems with frequent redistribution of the mesh whereas the graph
partitioning methods seem more favorable for productions runs on a fixed non-adaptive
grid.


\subsection{User Defined Partitioning}
\label{userdeflb}

A more general approach is provided by the methods
\begin{lstlisting}
  template< class LBDestinations >
  bool repartition ( LBDestinations &destinations );
  template< class LBDestinations, class DataHandleImpl, class Data >
  bool repartition ( LBDestinations &destinations, 
                     CommDataHandleIF< DataHandleImpl, Data > &dataHandle);
\end{lstlisting}
performing load balancing either without or with migrating user data using 
callback on the \code{dataHandle} instance.
Otherwise, the whole load balancing is taken care of by the user.
The class \code{LBDestinations} has to fulfill the following interface 
\begin{lstlisting}
  struct LBDestinations
  {
    // Return process number the given macro element should be assigned to.
    int operator()(const Grid::Codim<0>::Entity &);
    // Fill set of ranks the current process will receive elements from and return true
    // in this case. If false is returned, then ALUGrid will compute this information 
    // via a global communication.
    bool importRanks( std::set<int>& ranks ) const; 
  };
\end{lstlisting}
where the \code{int operator()(const Grid::Codim<0>::Entity &)} returns the process
number an element is to be moved to. In \dune[ALUGrid] this method will be
called for each macro element on the given rank and that macro element
together with all its children will be moved to the desired partition.
The method
\code{importRanks} can simply return \code{false} and then does not need to
fill the set \code{ranks}. However, this decreases performance
due to the global communication required to find out from which ranks to expect
data.
Some partitioning tools like \zoltan provide this information, so that the user only
needs to copy it to \code{ranks} vector and return \code{true} to improve parallel efficiency.

An example usage is shown in
\file{examples/loadbalancing/loadbalance_simple.hh}. The partitioning is
computed by keeping the center on process zero and distributing the rest of
the grid in equal slices to the other processors. 
The only changes required to the algorithm are in \file{main.cc} and
\file{adaptation.hh} where the calls of the \code{loadbalance(...)} method
on the grid are replaced with the new \code{repartition(...)} methods
In each step of the
scheme before calling \code{grid.repartition(...)} the method
\code{repartition()} is called on the loadbalance handle. This causes an 
internal variable to be increased, leading each time to a new partitioning:
\begin{lstlisting}
template< class Grid >
struct SimpleLoadBalanceHandle
{
  typedef SimpleLoadBalanceHandle This;
  typedef typename Grid :: Traits :: template Codim<0> :: Entity Element;
  SimpleLoadBalanceHandle ( const Grid &grid )
  : angle_( 0 )
  , maxRank_( grid.comm().size() )
  {}

  /** this method is called before invoking the repartition
      method on the grid, to check if the user-defined
      partitioning needs to be readjusted */
  bool repartition () 
  { 
    angle_ += 2.*M_PI/50.;
    return true;
  }

  /** This is the method, called from the grid for each macro element. 
      It returns the rank to which the element is to be moved. */
  int operator()( const Element &element ) const 
  { 
    typedef typename Element::Geometry::GlobalCoordinate Coordinate;
    Coordinate w = element.geometry().center();
    w -= Coordinate(0.5);
    if (w[0]*w[0]+w[1]*w[1] > 0.1 && maxRank_>0)
    { // distribute everything away from the center in equal slices
      double phi=arg(std::complex<double>(w[0],w[1]));
      if (w[1]<0) phi+=2.*M_PI;
      phi += angle_;
      phi *= double(maxRank_-1)/(2.*M_PI);
      int p = int(phi) % (maxRank_-1);
      return p+1;
    }
    else // keep the center on proc 0
      return 0;
  }

  /** This method can simply return false, in which case ALUGrid 
      will internally compute the required information through 
      some global communication. To avoid this overhead the user 
      can provide the ranks of partitions from which elements will
      be moved to the calling repartition. */
  bool importRanks( std::set<int> &ranks) const { return false; }
private:
  double angle_;
  int maxRank_;
};
\end{lstlisting}
A more useful example is given in
\file{examples/loadbalancing/loadbalance_zoltan.hh}, where the algorithm
in \dune[ALUGrid] relying on the \zoltan's graph partitioner is replicated
using the \dune interface. Note that the results will not be identical
since the order of the edges within the graph will differ slightly when
using the \dune interface to build it.
Nevertheless, the algorithm and parameter
settings for \zoltan are identical. Based on this implementation it is easy
to experiment with the wide range of options \zoltan provides to optimize
the partitioning algorithm for a given application. 
Note also that the class again contains a \code{repartitioning} method using the
same \code{lbOver}, \code{lbUnder} values provided in the \file{alugrid.cfg}
file. 

Constructing the graph relying only on the available \dune interface would
be quite cumbersome and involve quite a bit of overhead. There is no direct
way to compute the edge weights and the master rank for each ghost element
has to be passed on to \zoltan, information requiring an extra
communication step within \dune. To simplify constructing the graph
\dune[ALUGrid] provides a new method on the grid. 
\begin{lstlisting}
  template<PartitionIteratorType pitype>
  typename Partition<pitype>::MacroGridView macroView() const;
\end{lstlisting}
This method returns a view of the macro grid level of the grid. The
\code{MacroGridView} contains the usual method to iterate over the macro
grid and obtain an index set but in addition includes some useful 
methods to construct the dual weighted graph:
\begin{lstlisting}
  // return the master process of the given element
  int master ( const typename Codim< 0 > :: Entity &entity ) const;
  // return a globally uniqe integer id for this element
  int macroId ( const typename Codim< 0 > :: Entity &entity ) const;
  // return the weight (number of leaf elements) for the given elements
  int weight ( const typename Codim< 0 > :: Entity &entity ) const;
  // return the weight for this intersection 
  int weight ( const Intersection &intersection ) const;
\end{lstlisting}

The \zoltan example demonstrates a practical usage of the new load balancing 
interface and also an extension not directly available using the internal
bindings: The hypergraph algorithm of \zoltan can be
used to fix a set of elements to a given processor. By changing the
variable \code{fix_bnd_} to \code{true} the partitioning is computed such
that all elements adjacent to left boundary face are kept on process zero
throughout the simulation. A practical example of this possibility is
discussed in \cite{jehl:14}. It should be noted that, although the algorithm used in
this example mirrors the one used in the \dune[ALUGrid] internal bindings to \zoltan,
the results might not be the same. The reason is that iteration order over the macro
elements can differ and this results in slightly different dual graphs.

{\bf Note:} As pointed out above, \dune[ALUGrid] only allows to partition
the macro level of the grid. Depending on the problem the macro grid might
not contain enough elements or the adaptivity might be too localized to
allow for a balanced load if only macro elements are distributed. On
manycore systems a possible solution is to use fewer processes to distribute the
macro grid and use threading to partition directly on
the leaf level. This approach has been evaluated in 
\cite{dgimpl:12}.



\section{Conclusions}

In this paper we briefly described the main new features available in the overhaul
of \dune[ALUGrid]. The main improvements concern the
parallel feature set of the library, including now user-defined load
balancing and parallel grid construction as well as a decreased memory footprint.
Since \alugrid is and was widely used within the \dune community we expect that
numerous \dune users will benefit from work presented here. 
We also presented a number of extensions to the \dune grid interface that prove useful 
and will hopefully be integrated into the \dune grid interface in the near future.

The increased feature set also includes newest vertex bisection for tetrahedral
grids in 3d, making it the only parallel grid manager within \dune with
this feature. 
This will enable the usage of conforming adaptive discretization methods, such as
conforming adaptive Finite Elements, in parallel. Nevertheless, there are some
shortcomings that still have to be resolved in the future.

The 2d code has been parallelized by reformulating it as an extension to the 3d code and it thus also inherited all the major features. The usage is similar to the usage of the 3d code. 

\subsection{Shortcomings and Outlook}

A major drawback of the current implementation is that load balancing
is performed solely based on the
macro grid. This works fine for many problems where the refinement zones are
not too restricted to one area of the domain, but will completely fail for
very local refinement regions. As already mentioned the situation can be
improved by using a hybrid parallelization approach. But the next major improvement will be the
implementation of a more flexible partitioning of elements allowing for partitioning of 
various sets of elements. Furthermore, the current implementation 
lacks support for ghost elements when bisection refinement is used. This is
hopefully fixed in the near future. 

\section*{Acknowledgements}
We would like to acknowledge high-performance computing
support from Yellowstone \cite{Yellowstone} provided
by NCAR's Computational and Information Systems
Laboratory, sponsored by the National Science Foundation.
Robert Kl\"ofkorn acknowledges
the DOE BER Program under the award DE-SC0006959 and the 
National IOR Centre of Norway.



\end{document}